\documentclass[aip,jcp,numerical,preprint,titlepage,showkeys,nofootinbib,nobibnotes,superscriptaddress]{revtex4-1}

\usepackage{graphicx}
\usepackage{amssymb,amsfonts,amsmath}
\usepackage{amsfonts}
\usepackage{amsmath}
\usepackage{calc}
\usepackage{longtable}
\usepackage{tabularx}
\usepackage{mathrsfs}
\usepackage{array}
\usepackage{url}
\usepackage[sort&compress]{natbib}

\date{20180813}

\begin{document}
\title{Aerosol-OT surfactant forms stable reverse micelles in aploar solvent in the absence of water} 

\author{Ryo Urano}
\altaffiliation[current addrss: ]{Engineering Department, Nagoya University}  
\affiliation{ Chemistry Department, Boston University, 590 Commonwealth Avenue, Boston, MA 02215 USA}

\author{George A. Pantelopulos}
\affiliation{ Chemistry Department, Boston University, 590 Commonwealth Avenue, Boston, MA 02215 USA}

\author{John E. Straub}
\affiliation{ Chemistry Department, Boston University, 590 Commonwealth Avenue, Boston, MA 02215 USA}
\altaffiliation[corresponding author e-mail:]{straub@bu.edu }  

 \keywords{reverse micelle, size distribution, energy representation, molecular dynamics }

\begin{abstract}

Normal micelle aggregates of amphiphilic surfactant in aqueous solvent are formed by a process of entropically driven self-assembly.  The self-assembly of reverse micelles from amphiphilic surfactant in non-polar solvent in the presence of water is considered to be an enthalpically driven process.  While the formation of normal and reverse surfactant micelles has been well characterized in theory and experiment, the nature of dry micelle formation, from amphiphilic surfactant in non-polar solvent in the absence of water, is poorly understood.  In this study, a theory of dry reverse micelle formation is developed. Variation in free energy during micelle assembly is derived for the specific case of AOT surfactant in isooctane solvent using atomistic molecular dynamics simulation analyzed using the energy representation method.  The existence and thermodynamic stability of dry reverse micelles of limited size are confirmed.  The abrupt occurence of monodisperse aggregates is a clear signature a critical micelle concentration, commonly observed in the formation of normal surfactant micelles.  The morphology of large dry micelles provides insight into the nature of the thermodynamic driving forces stabilizing the formation of the surfactant aggregates.  Overall, this study provides detailed insight into the structure and stability of dry reverse micelles assembly in non-polar solvent.

\end{abstract}

\maketitle

\section*{Introduction}
\label{sec:orgheadline1}

The rich phase behavior and complex dynamics of surfactant microemulsions have been a focus of intense experimental, theoretical, and computational study for decades.  The detailed dynamic and thermodynamic behavior of ternary mixtures of surfactant, oil, and water are fundamentally important to the theory of complex solutions.   In addition, self-assembled structures such as micelles, reverse micelles, and membranes have great applied importance to biology, as well as environmental and industrial chemistry.   As such, the development of a fundamental understanding of the equilibrium state of microemulsions has been a critical goal for the field.  

The reverse micelle (RM) is a phase of particular interest, in which surfactant aggregates containing a water core are suspended in non-polar solvent.  The RM morphology has been exploited for a variety of applications, including chemical synthesis\cite{zhang_characterization_2007}, 
drug delivery systems\cite{hino_basic_2000,okochi_preparation_2000,higashi_hepatic_2000}, 
studies of model membranes\cite{nishii_transport_2002}, 
and solute encapsulation\cite{nucci_high-resolution_2014,dodevski_optimized_2014}. 
While an empirical approach to the optimization of surfactant mixtures has led to significant advancement, it has proven difficult to physically characterize RM solutions in terms of the distribution of aggregate size and nature of RM structure.  As such,  there is a pressing need to develop a first-principles theory for the {\it de novo} prediction of the RM size distribution as a function of solution composition.

The assembly of normal surfactant micelle in water solvent has long been assumed to be driven by an increase in water entropy following surfactant aggregation and exclusion of water from the micelle interior, leading to favorable changes in entropy and enthalpy upon micelle assembly\cite{galamba_waters_2013,galamba_water_2014,schick1987nonionic,aranow_environmental_1960,aranow_additional_1965,shiraga_hydration_2014,frank_free_1945}.  
The assembly of ``wet'' RMs from mixtures of surfactants, oils, and water in ambient conditions is typically considered to be enthalpy driven, with the RM phase stabilized by favorable interaction of water and surfactant head groups\cite{Eicke1978}, with size distributions determined by water loading\cite{VanDijk1989} and salt content\cite{Fathi2012} attributed to electrostatic interactions\cite{chowdhary_molecular_2009, Eskici2016}. The unique aqueous environment experienced by molecules encapsulated in RMs\cite{faeder_molecular_2000, faeder_solvation_2001, chowdhary_molecular_2011} has been utilized in molecular synthesis\cite{Graeve2013} and the study of protein structure\cite{Mukherjee2006, Mukherjee2007}.
In contrast, clear identification of the principal driving force underlying the formation of ``dry'' RMs from surfactant in oil solvent in the absence of water has remained elusive\cite{mukherjee1993thermodynamics}. As such, the mechanism of surfactant aggregation in oil solvent in the absence of water and the very existence of dry RMs (hereafter referred to as dRMs) continues to be debated.

In early theoretical work, Ruckenstein and Nagarajan\cite{ruckenstein1980aggregation} used a free energy functional approach for AOT surfactant in non-polar solvent to argue that interactionf between surfactant head groups and tails creates a free energy minimum associated with the stable formation of dRMs.  Stable dRM aggregates were predicted to be restricted to a gradual increase in ``aggregation number'' (number of AOT surfactant molecules in a given micelle) of less than ten, suggesting the absence of a critical micelle concentration (CMC).  Given this prediction, it has been suggested that the experimentally observed CMC in dRM mixtures must result from the presence of trace water molecules (carried over from AOT synthesis and incomplete surfactant ``drying'') in a ratio of less than one water per AOT molecule. 
Motivated by experiments\cite{ekwall1970some} suggesting an inverse hexagonal structure in pure sodium AOT solutions, Harrowell et al. \cite{Wootton2008} investigated the structure of dRMs in sodium-sulfate ion clusters in the vacuum state.  Considering the observed sodium-sulfate ion cluster structure and the structure of AOT reverse micelles as a perturbation of the crystal structure, they obtained a mean aggregation number that was macroscopically large, suggesting that AOT is insoluble in oil solvent. 

A variety of experimental studies have confirmed the existence of dry AOT aggregates.  Calorimetric studies\cite{TANAKA1992,Tanaka2005,Sameshima2006} confirmed that the stabilizing interaction energy between AOT molecules in organic solvents is so low as to make the formation of AOT oligomers improbable. In contrast, studies based on neutron scattering\cite{Chen1986} and absorption spectroscopy\cite{Muto1973} support the existence of a critical micelle concentration.  In addition, a variety of studies have led to independent assessments of mean dRM size (the mean number, $\bar{n}$, of surfactants in dRMs), including $\bar{n}$=18 by vapor pressure\cite{Kon-no1971}, and $\bar{n}$=39 and $\bar{n}$=56 by light scattering \cite{Kitahara1962,Frank1969}. More recent small-angle neutron scattering studies\cite{Smith2013,smith2016effect} have shown a sharp transition in the mean dRM size as a function of AOT concentration in non-polar solvents.  Taken together, these experimental studies provide clear support for the existence of a CMC related to dRM formation at higher AOT concentrations.  Interestingly, these studies also demonstrate that the dRM size distribution sensitively depends on solvent properties, such as dielectric constant and surfactant solubility\cite{hollamby_effect_2008}.

Significant theoretical work has been done to define the size distribution of surfactant micelles in water solvent.   Christopher and Oxtoby\cite{christopher2003free} proposed a density functional model that assessed free energy contributions to the aggregation process, obtaining a size distribution from numerical minimization of the free energy.  Mohan and Kopelevich\cite{mohan2008multiscale} obtained kinetic rate constant for formation of spherical micelles formed by non-ionic surfactants using a coarse grained model and kinetic analysis.   Kindt and coworkers\cite{Kindt2013} employed the chemical species model\cite{Ben-Shaul1982} and a statistical dynamical equation for the size distribution parameterized with an equilibrium constant obtained from molecular simulation. Kinoshita and Sugai\cite{kinoshita1999new,kinoshita2002methodology} also employed the chemical species model but used the reference interaction site model (RISM) theory in their simulations to obtain the chemical potential for surfactant aggregates of varying size. Additionally, Kindt and coworkers have expanded on this work to create the partition-enabled analysis of cluster histograms (PEACH) method via novel applicaiton of number theory to evaluation of changes in partition function in the chemical species model, enabling rapid and precise refinement of measured reaction rates between $n$-mers in solutions at equilibrium.  In an impressive theoretical study of the equilibrium properties of micelle formation, Yoshii, Okazaki, and others \cite{Yoshii2006,Yoshii2006a,Yoshii2006b,Yoshii} employed the chemical species model to obtain a thermodynamic relation defining micelle size in terms of the free energy of surfactant insertions. Their work employed thermodynamic integration at infinite dilution and surfactant activity coefficients proposed via fitting of experimental data using Debye-H{\"u}ckel theory, resulting in a {\it de novo} theoretical prediction of the CMC and distribution of micelle size as a function of surfactant concentration.

In this work, we extend the seminal work of Yoshii and Okazaki to study the size distribution of aggregates of AOT in pure isooctane solvent.  In order to address the inherent complexity of the dRM system, the chemical potential for growth of dRMs is evaluated using the classical free energy functional based Energy Representation (ER) method\cite{Matubayasi2000,Matubayasi2002,Matubayasi2003}.  The ER method is shown to dramatically increase the efficiency of free energy calculation within sufficient accuracy to provide thermodynamically relevant results.  The resulting theory provides a {\it de novo} prediction of the surfactant aggregate size distribution, providing evidence for the existence of stable dry AOT RMs and surprising insight into the RM structure and stability.

This paper is organized as follows. In the Methods section, we derive a thermodynamic relation for the relative stability of AOT aggregates as a function of aggregate size.  The theoretical foundation for the evaluation of the chemical potential as a function of aggregate size based on free energy calculations is described in detail.  In the Results section, 
we first determine the free energy of dRM growth by addition of solvated surfactant monomers in the energy representation method.  Subsequently, the structures and relative free energies of AOT dRMs are characterized as a function of aggregation size.  The mean size, $\bar{n}$, is derived and interpreted in terms of size-dependent aggregate structural changes.  Finally, the theoretical results are used to address discrepancies between prior theoretical predictions and experimental observations.  Overall, this study provides the first detailed atomic-level characterization of the equilibrium distribution of dry AOT reverse micelles.

\section*{Methods}
\label{sec:orgheadline6}

\subsection*{Thermodynamic relation for size distribution of each composition}
\label{sec:orgheadline2}

The theoretical foundation of our approach is a definition relating the relative concentration of a dry reverse micelles (dRM) of a given size to the free energy difference between dRMs of varying size and the corresponding activity coefficient. The starting point is the previous works of Okazaki and coworkers exploring the thermodynamic properties of surfactant micelles\cite{Yoshii2006b,Yoshii2006,Yoshii2006a,Yoshii}.

The chemical species model is employed\cite{tanford1980hydrophobic,puvvada1990molecular,zoeller1997statistical,christopher2003free,Kindt2013,kinoshita1999new} in which the composition of the solution is defined in terms of a distribution of dRMs of varying size assuming each size as a distinct species. The chemical potential $\mu_{n}$ of an aggregate composed of $n$ surfactant monomers, $nA$, is written
\begin{equation}
  \mu_{n} = \mu_{n}^0 + k_B T \ln{ a_{n}}, \label{eq:0}
\end{equation}
where $\mu_{n}^0$ is the chemical potential of a dRM of aggregation number $n$ at standard state and $a_{n}$ is the relative activity of a micelle of aggregation number $n$ in solution at a certain dilution. The standard state is a solvated monomer at infinite dilution, as it is often defined for liquid phase solutes.

The free energy change accompanied by the formation of a dRM of $n$ solutes in solution, $R_{n}^{\text{(sol)}}$, from $n$ isolated surfactant molecules in vacuum and a pure solvent system is described by

\begin{equation}
  n\ A^{\textrm{ (vac) }} \overset{{
  \mu_{n}^{0}}}{\rightleftharpoons}   R_{n}^{(\textrm{sol})} \label{eq:3}.
\end{equation}

For solvent molecules, the chemical potential, $ \mu _s$, is defined as
  \begin{equation}
   \mu _s = \mu _s ^0 + k_B T \ln{a_s}
  \end{equation}
in terms of the standard state chemical potential $\mu _s ^0$ of the solvent and the corresponding activity coefficient, $a_s$.
The total free energy of a solution containing dRMs of various sizes, $G$, can be written

\begin{eqnarray}
  G &=& N_{s} \mu_{s} + \sum_{n} N_{n} \mu _{n}   \\
   &=& 
N_{s} \mu _{s}^{0} +  \sum_{n}  \mu _{n}^{0} 
+ k_{B}T (N_{s} \ln{a_{s}} 
 +  \sum_{n} N_{n} \ln{a_{n}}
  ),  
\end{eqnarray}

\noindent where $N_n$ is the number of each size of dRM, and $N_s$ is the total number of solvent molecules. 

For the association reaction adding one amphiphilic surfactant molecule to an aggregate of $n$ surfactant molecules in solution
  \begin{eqnarray}
      R_{n}^{\textrm{ (sol)}}  + \textrm{A}^{\textrm{ (sol)}}
 \overset{{\Delta G}_{n+1}}{\rightleftharpoons}  R_{n+1}^{\textrm{ (sol)}}     \label{eq:6}
  \end{eqnarray}
the change in free energy is the difference in chemical potentials for insertion of $n$$+$1 from $n$ and $1$ monomers from vacuum to form dRMs  described in Eq. (\ref{eq:3}). We decompose these chemical potentials to standard state contributions, $\mu^{0}$, and contributions from the activity (describing \emph{extra}-micellar interactions), $\mu^{a}$, as
          \begin{eqnarray}
            \Delta G_{n+1} &=& \mu _{n+1} - (\mu_{n} + \mu _{1} )  \\
            &=& (\mu_{n+1}^{0} + \mu_{n+1}^{a}) - (\mu_{n}^{0} + \mu _{n}^{a}) - (\mu_{1}^{0} + \mu _{1}^{a})  \\
            &=&  {\Delta \mu}_{n+1}^{0}  
             + k_BT \ln { \frac{X_{n+1}}{X_{n} X_{1} }} + k_BT \ln{ \frac{\gamma _{n+1} }{\gamma _{n} \gamma _{1} }} ,
          \end{eqnarray}
where the difference in chemical potential for adding one surfactant to a preexisting dRM of aggregation number $n$ at standard state is
           \begin{eqnarray}
      {\Delta \mu}_{n+1}^{0} &\equiv& \mu _{n+1}^0 - (\mu_{n}^0 + \mu_{1}^0 ). \label{eq:5}
          \end{eqnarray}

The activity coefficient $a_{n}$ is defined in terms of the mole fraction $X_n$ and activity coefficient $\gamma_n$ at size $n$ as $a_{n} = \gamma _{n} X_{n}$.  At equilibrium $\Delta G_{n+1} = 0$, for which we obtain the final relation defining the dRM size distribution
\begin{equation}
     \frac{X_{n+1}}{X_{n}} =\frac{\gamma _{n} \gamma _{1} }{\gamma _{n+1} } X_{1}  \exp{\left(    \frac{  -{\Delta \mu}_{n+1}^{0}  }{k_BT}\right)}\label{eq:1} 
\end{equation}
where $X_n$ is the mole fraction of reverse micelles of aggregation number $n$, and $\Delta {\mu}_{n+1}^{0}$  is the change in chemical potential associated with the reaction in Eq. (\ref{eq:6}) taken to be at standard state (infinite dilution).

Additionally, the effect of molecular indistinguishability may need to be accounted for in practice, depending on the method used to determine $\Delta G_{n+1}$. Kindt has proposed a correction to account for molecular distinguishability\cite{kindtpersonal}.  For the reaction described by $\Delta G_{n+1}$
  \begin{eqnarray}
      R_{n}^{\textrm{ (sol)}} + \textrm{A}^{\textrm{ (sol)}}
 {\underset{k_2}{\stackrel{k_1}{\rightleftharpoons}}}  R_{n+1}^{\textrm{ (sol)}}, \label{eq:kindt1}
  \end{eqnarray}
the forward rate, $k_1$, involves only one molecule, $A$. However, the backward rate, $k_2$, involves the ($n+1$) indistiguishable members of $R_{n+1}^{\text{ (sol)}}$. From $R_{n+1}^{\text{ (sol)}}$ one of the $n+1$ indistinguishable AOTs will disassociate, leaving an AOT $n$-mer. Accounting for this will slightly change the free energy such that

  \begin{eqnarray}
     \Delta G_{n+1}^{corrected} = \Delta G_{n+1} + k_\text{B}T \ln \left( n+1 \right). \label{eq:kindt2}
  \end{eqnarray}

This correction scales as $\ln(n)$ and may be considered negligible in many cases when thinking of free energy changes over small ranges of $n$ beyond $n$=10. As such, this correction does not change the qualitative results of our study or the past work of Yoshii and coworkers.\cite{Yoshii2006,Yoshii2006a,Yoshii2006b,Yoshii} Careful consideration of such reaction rates may need to be considered depending on the study and methodology employed in future works.

\subsection*{Physical meaning of thermodynamic quantities}
\label{sec:orgheadline3}

To obtain the size distribution of RMs according to Eq. (\ref{eq:1}), it is necessary to determine the aforementioned chemical potential difference at infinite dilution \( {\Delta \mu}_{n+1}^{0}\) and activity coefficient \(\gamma_n\) for all physically relevant values of $n$.  The activity coefficient \(\gamma_{n}\) is often derived from fitting experimental solvation data to the predictions of the Debye-H{\"u}ckel theory.

Consider the insertion of a surfactant molecule into a dRM of given aggregation number $n$ leading to the formation of a dRM of aggregation number $n$$+$1.  The ratio of concentrations  of the two species can be related to the change in free energy upon surfactant molecule insertion as

\begin{equation}
  \frac{X_{n}}{X_{1}} = \left(\frac{\gamma _{1}}{\gamma _{n}}
 \right) 
\left( \gamma _{1} X_{1} \right)^{n-1} 
\exp{\left( \frac{ \left[ \sum_{i=2}^{n} - {\Delta \mu} _{i}^{0}  
\right]
}{k_{B} T} \right)  } \label{eq:11}
{\rm ~ for}~ n\geq 2.
\end{equation}

Having computed the dRM size distribution in terms of the concentrations of dRM sizes,  the law of mass action can be used to relate the relative concentrations of dRMs to the solute concentration of surfactant as

\begin{equation}
 \sum_{n} n X_{n} = N_{AOT}/N_{ISO}.\label{eq:massaction}
\end{equation}
\subsection*{Free energy evaluation}
\label{sec:orgheadline4}

In order to determine the free energy change upon insertion of a surfactant molecule into a preexisting dRM we employ the energy representation (ER)\cite{Matubayasi2000,Matubayasi2002,Matubayasi2003}.
We briefly introduce the formally exact free energy evaluation method of thermodynamic integration (TI) followed by a discussion of the approximate energy representation method (ER).  

Following the notation of Frolov\cite{Frolov2015}, the potential energy function of the solute-solvent system is written
\begin{eqnarray}
  V(\mathbf{ r}_s, \mathbf{ r}_w  ) = \Phi (\mathbf{ r}_w ) + v(\mathbf{ r}_s, \mathbf{ r}_w )
\end{eqnarray}
consisting of the solvent-solvent potential $\Phi (\mathbf{ r}_w )$ and full coupling solute-solvent potential $v(\mathbf{ r}_s, \mathbf{ r}_w )$, where $\mathbf{r}_s$ represents the configuration of the solute molecule and $\mathbf{r}_w$ the configuration of the solvent molecules. 
 We assume the $\lambda$-dependent solute-solvent interaction is linearly modulated by parameter $\lambda$  as
\begin{eqnarray}
  V(\mathbf{ r}_s, \mathbf{ r}_w; \lambda) = \Phi (\mathbf{ r}_w ) + u_\lambda(\mathbf{ r}_s, \mathbf{ r}_w )= \Phi (\mathbf{ r}_w ) + \lambda v(\mathbf{ r}_s, \mathbf{ r}_w ) 
\label{eq:int}
\end{eqnarray}
where $\lambda=0$ represents the pure solvent system  $V(\mathbf{ r}_s, \mathbf{ r}_w; \lambda=0 ) = \Phi (\mathbf{ r}_w )$ and $\lambda=1$ represents the full interaction between solute and solvent  $V(\mathbf{ r}_s, \mathbf{ r}_w; \lambda=1 ) = \Phi (\mathbf{ r}_w ) + v(\mathbf{ r}_s, \mathbf{ r}_w )$.

\medskip
\noindent{\sl {Thermodynamic Integration (TI) approach}}
\medskip

The excess chemical potential (change in solvation free energy) of the solute can be written using Kirkwood's charging formula 
  \begin{eqnarray}
           \mu^{ex} &=& G_{u_{\lambda=1} }(\mathbf{ r}_s,\mathbf{ r}_w )-G_{u_{\lambda=0} }(\mathbf{ r}_s,\mathbf{ r}_w )  \nonumber \\
    & =& \int_{ 0}^{1} d \lambda \int d\mathbf{ r}_s d\mathbf{ r}_w 
          \frac{\partial u _{\lambda} (\mathbf{ r}_s,\mathbf{ r}_w )}{\partial  \lambda }
      \rho_{\lambda} (\mathbf{ r}_s, \mathbf{ r}_w ) =  \Large\langle  \frac{\partial u _{\lambda} (\mathbf{ r}_s, \mathbf{ r}_w )}{\partial  \lambda }\Large\rangle  _{\lambda}
\label{eq:TI}
  \end{eqnarray}
  which forms a popular foundation for thermodynamic integration (TI).  The quantity $  \rho_{\lambda} (\mathbf{ r}_s, \mathbf{ r}_w )$ is the normalized classical density distribution corresponding to a potential energy of interaction given by Eq. (\ref{eq:int}) for a particular value of $\lambda$.  In TI,  knowledge of intermediate states between $\lambda=0$ and $\lambda =1$ is required for evaluation of the $\lambda$ integral.  As such, the ensemble average of $\partial u _{\lambda}/\partial \lambda$  is required for each value of $\lambda$.
  An effective parameterization of $V(\mathbf{ r}_s, \mathbf{ r}_w; \lambda)$ and sufficient sampling at intermediate and end states are essential to the success of the TI approach.

\medskip
\noindent{\sl {Energy Representation (ER) method}}
\medskip

 An alternative to the formally exact TI approach is the approximate energy representation (ER) method. In the ER method, integration over configuration space of the solute and solvent is replaced by integration over the interaction energy between solute and solvent.   The classical density distribution $\rho_{\lambda} (\mathbf{ r}_s, \mathbf{ r}_w )$ is replaced by the probability density of specific values of the interaction potential
\begin{equation}
   \rho^{}_{sw,\lambda}(\epsilon) = \sum_{i=1}^{N_w} \delta ( v(\mathbf{ r}_s, \mathbf{ r}_w  ) - \epsilon ) 
\end{equation} 
where the energy coordinate is defined as  $u_{\lambda}^{}(\epsilon)=\int_{-\infty}^{\infty} d {\bf r}_{s} d {\bf r}_{w} \delta (v ({\bf r_s}, {\bf r}_w  ) - \epsilon )  u_{\lambda}  ({\bf r_s}, {\bf r}_w  )$.   

The completeness and equivalence of the energy representation to the phase space representation are supported by the Kohn-Sham density functional theorem. 
In the ER method, information for parameterizing the energy density is obtained from computer simulation.

The formally exact result for $\mu^{ex}$ given by Eq. (\ref{eq:TI}) may be reformed as
  \begin{eqnarray}
  \mu^{ex} & = & \int_{ 0}^{1} d \lambda \int d\mathbf{ r}_s d\mathbf{ r}_w 
          \frac{\partial u _{\lambda} (\mathbf{ r}_s, \mathbf{ r}_w )}{\partial  \lambda }
      \rho_{\lambda} ( \mathbf{ r}_s, \mathbf{ r}_w )   \nonumber 
\label{eq:ER} \\
   &=& \int_{- \infty}^{\infty}  d \epsilon   \rho^{}_{sw,\lambda=1} (\epsilon ) \epsilon - \int_{0}^{1} d\lambda 
   \int_{-\infty}^{\infty} d\epsilon \frac{\partial  \rho^{}_{sw,\lambda} (\epsilon)}{\partial \lambda }  u^{}_{sw,\lambda} (\epsilon ),
  \end{eqnarray}
  where  $u^{}_{sw,\lambda=1} (\epsilon )=v_{sw}^{}(\epsilon)=\epsilon$ from the definition of the energy coordinate.  The first term corresponds to the contribution of the solute-solvent self energy. 
  To evaluate the second term, we introduce an auxiliary function $\omega ^{}_{sw,\lambda}$, which is the analogue of the the potential of mean force and defined through the relation
  \begin{equation}
    \rho^{}_{sw,\lambda}(\epsilon ) = \rho ^{} _{sw,\lambda=0} (\epsilon ) \exp{\left[ -\beta ( u^{}_{sw,\lambda } (\epsilon )
  + \omega ^{}_{sw,\lambda} (\epsilon )
  )\right]}.
  \end{equation}
  The function  $\omega ^{}_{sw,\lambda} (\epsilon)$ captures the many-body interaction between solute and solvent.  

 The energy density is written as the linear combination
 \begin{equation}
  \rho^{}_{sw,\lambda} (\epsilon) = \lambda \rho^{}_{sw,\lambda=1} (\epsilon)+ (1-\lambda )\rho^{}_{sw,\lambda=0} (\epsilon)
 \end{equation}

\noindent and given the potential of mean force and direct interaction energy $  u^{}_{sw,\lambda} (\epsilon )$, we can transform the second term in Eq. (\ref{eq:ER}) as
  \begin{eqnarray}
  \mathscr{F} \left[ \rho^{}_{sw,\lambda} (\epsilon) ,  u^{}_{sw,\lambda} (\epsilon )
  \right]  & \equiv & - \int_{0}^{1} d\lambda 
   \int_{-\infty}^{\infty} d\epsilon \frac{\partial  \rho^{}_{sw,\lambda} (\epsilon)}{\partial \lambda }  u^{}_{sw,\lambda} (\epsilon ) \\
  &=& k_{B} T \int_{  } d \epsilon \left[
        ( \rho^{}_{sw,\lambda=1} (\epsilon )  - \rho^{}_{sw,\lambda=0} (\epsilon ))   \nonumber
        - \rho^{}_{sw,\lambda=1} (\epsilon) \ln{\frac{\rho^{}_{sw,\lambda=1} (\epsilon ) }{\rho^{}_{sw,\lambda=0} (\epsilon ) }}
\right. \\ 
       && \left.  -\beta ( \rho^{}_{sw,\lambda=1} (\epsilon )  - \rho^{}_{sw,\lambda=0} (\epsilon ))  \int_{0}^{1} d \lambda \omega _{sw,\lambda}(\epsilon)   \right],
    \end{eqnarray}
    where we have used the identity
\begin{equation}  
  \frac{\partial  \rho^{}_{sw,\lambda} (\epsilon)}{\partial \lambda } =  \rho^{}_{sw,\lambda=1} (\epsilon )  - \rho^{}_{sw,\lambda=0} (\epsilon ).
\end{equation}

\noindent Finally, the excess chemical potential  in the energy representation can be written
\begin{equation}
    \mu^{ex} \left[ \rho^{}_{sw,\lambda} (\epsilon),  u^{}_{sw,\lambda} (\epsilon )\right] =
  \int_{- \infty}^{\infty} d \epsilon \rho^{}_{sw,\lambda=1} (\epsilon) \epsilon - \mathscr{F} \left[ \rho^{}_{sw,\lambda} (\epsilon),  u^{}_{sw,\lambda} (\epsilon )\right]
\end{equation}

\noindent This equation is exact if the potential mean force $\omega^{}_{sw,\lambda} (\epsilon)$ is exact. However, in practice as in all integral equation theories $\omega ^{}_{sw,\lambda}$ must be treated approximately.
As such, the accuracy of this method depends on how well the approximate form of the potential of mean force captures many body interactions that are not included in the direct interaction of solute and solvent.

The pioneering work of Matubayasi and Nakahara provides guidance on the best choice of functional forms for the potential of mean force $\omega ^{} (\epsilon)$.  It is recommended that a combination of (1) a hypernetted chain (HNC) equation inspired contribution for $\omega^{(HNC)} (\epsilon) < 0$, capturing attractions, is combined with (2) a Percus-Yevick (PY) equation inspired term for $\omega ^{(PY)} (\epsilon) > 0$, capturing repulsions.  The sign of $\omega(\epsilon )$ shows the clear boundary between the repulsions and attractions. By exploiting this insight, the $\lambda$ integral can be heuristically weighted through a function $\alpha (\epsilon)$ as
 \begin{eqnarray}
    \beta  \int_{0}^{1} d \lambda   \omega^{}_{sw,\lambda} (\epsilon)   
  \approx   \alpha (\epsilon ) F_{\omega_1} (\epsilon) +  \left( 1-\alpha (\epsilon)\right) F_{\omega _0}(\epsilon),
  \end{eqnarray}
where $     F_{\omega_1}(\epsilon)= \beta \int_{0}^{1} d \lambda   \omega^{}_{sw,\lambda=1} (\epsilon) $ and  $    F_{\omega_0} (\epsilon ) =   \beta \int_{0}^{1} d \lambda   \omega^{}_{sw,\lambda=0} (\epsilon)  $. The specific form of  $\alpha(\epsilon)$ is discussed in detail elsewhere\cite{Matubayasi2002,Frolov2015}.   In this way, the $\lambda$ integral required for the determination of the excess chemical potential in the ER method may be evaluated solely based on knowledge of the energy density at the endpoints $\lambda=1$ and $\lambda =0$ through $\rho^{}_{sw,\lambda=1} (\epsilon)$ and  $\rho^{}_{sw,\lambda=0} (\epsilon)$.  In this way, the simulation of intermediate states for the system is avoided, providing a distinct advantage over the TI method.  However, many-body effects are approximately included. In systems for which many body interactions are important, the accuracy of this method is expected to diminish.

\medskip
\noindent{\sl {Application of the ER method}}
\medskip

We employed the ER method to determine the chemical potential differences (Eq. \ref{eq:5}) to understand the size distribution of AOT dRMs in pure isooctane solvent within a thermodynamically relevant range of $n$.  We chose to investigate dRMs of $n$ = 1, 10, 20, 30, 40, 50, 60, 70, 75, 80, 90, and 100, requiring simulation of each corresponding $n$$+$1 dRM for each $n$ dRM.  Rather than directly compute the formation of $n$-mers from vacuum, which we defined to derive Eq. (\ref{eq:1}), we directly evaluated $\Delta \mu^{0}_{n+1}$ by using the ER method to calculate the free energy of insertion of monomer in solution to a $n$-mer in solution.  To calculate $\Delta \mu^{0}_{n+1}$, we used the ER method via the ERmod program (version 0.3.4) \cite{Sakuraba2014}.  

\begin{figure}[htb]
\centering
\includegraphics[width=0.45\linewidth]{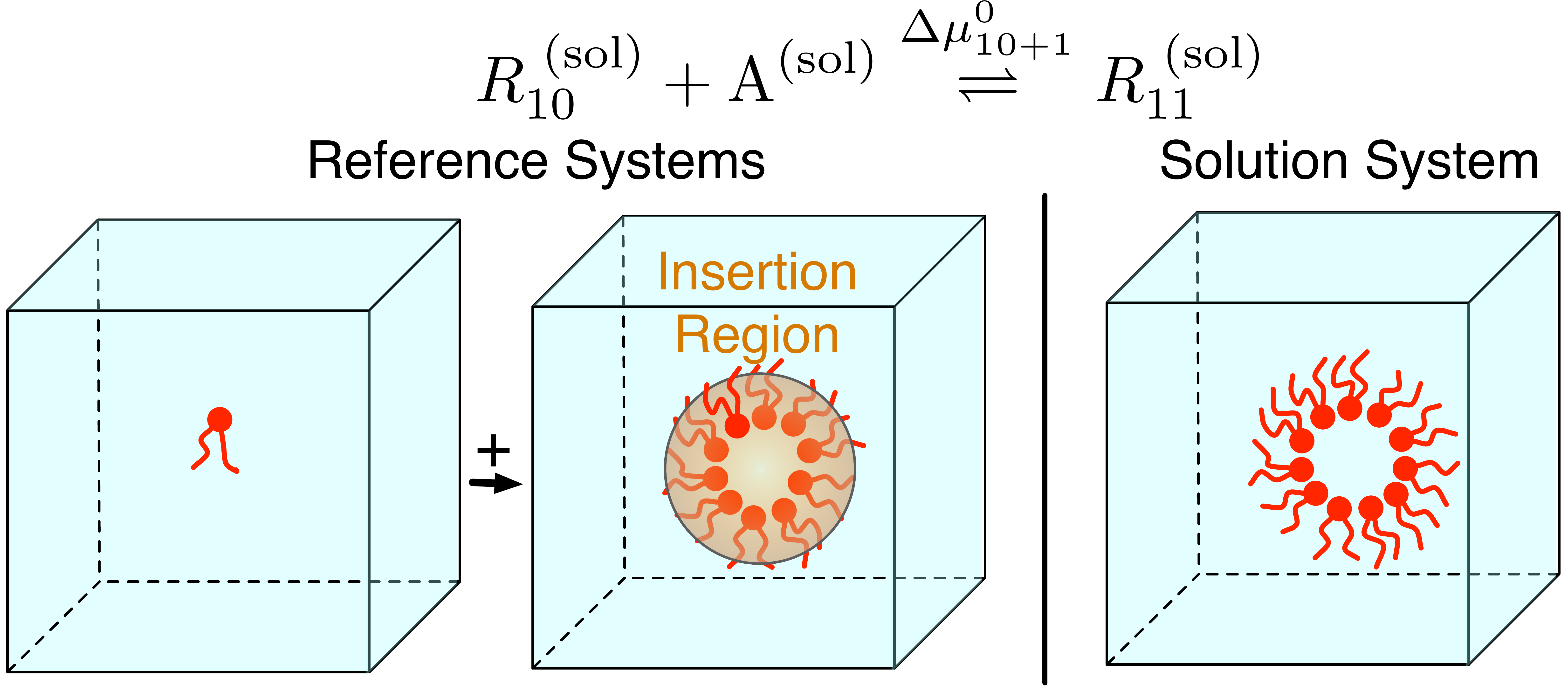}
\caption{\label{ERmethod}
Systems involved in determination of the free energy of AOT insertion to solvent from vacuum and AOT association of a monomer from solution to a 10-mer. The orange region indicates the ensemble-averaged radius of gyration of the 10-mer, to which solution monomer test insertion centers of mass were constrained.}
\end{figure}

For example, $\Delta \mu_{10}^{0}$ was calculated by evaluation of $\Delta \mu^{0}$ for insertion of an AOT monomer from a dilute solution to a dilute solution of isooctane solvent containing a dRM of 10 AOTs, as illustrated in Fig. \ref{ERmethod}.  This was accomplished by evaluating the interaction density function via simulation of (1) the ``solution'' system of an AOT 11-mer in isooctane solvent and (2) the ``reference'' systems of two independent simulations, one of an AOT 10-mer in isooctane and one of a simulation of an AOT monomer in isooctane. The reference systems were used to evaluate interaction energies of AOT monomer test insertions to the 10-mer solution. Test insertions of AOT monomer centers of mass were constrained to a spherical region described by the radius of gyration ($R_g$) of the $10$-mer dRM, $0$$<$$r$$<$$R_g$. Fig. \ref{ERmethod} describes the simulations needed to determine $\Delta \mu^{0}_{10+1}$.

Test insertions of the solute were performed 10,000 times for each reference configuration. Error bars were computed using the following scheme.  (1) Averages were calculated for the solution system with statistical error measured by splitting the trajectory into ten-block subtrajectories.  (2) Energy distributions were constructed in which the energy was discretized, respecting a mesh size of $\Delta \epsilon$.  The mesh size was increased from a chosen size to that value multiplied by a factor of 1, 2, 3, ..., and 10, e.g. $\Delta \epsilon$=0.001, 0.002, ..., and 0.01 kcal/mol.  $\Delta \mu^{0}$ computed for each mesh size was used as a measure of the mesh size error. Differences in $\Delta \mu^{0}$ between trajectory blocks using the same mesh size were found to be at most 2.0 kcal/mol for all aggregate sizes. As an example, the composite error of roughly 12 kcal/mol for $n=90$ is composed of the statistical error in the trajectory average of 2.0 kcal/mol and the error associated with mesh size dependence of 10 kcal/mol.

\subsection*{System setup}
\label{sec:orgheadline5}

Initial conditions were defined by spherically arranged configurations of surfactant molecules
created using Packmol\cite{martinez2009packmol}.  Isooctane solvent molecules were arranged surrounding the sphere of AOT, leaving a vacuum in the dRM center. The energy of each system was minimized using the steepest descent algorithm to remove bad contacts. Each
system was subsequently equilibrated in the NVT ensemble with velocity rescaling for 300 ps, followed by equilibration in the NPT
ensemble with Nose-Hoover and Parrinello-Rahman coupling for 300 ps.  Finally, each system
was equilibrated at 300 K and 1 bar to find an average density of approximately 0.7 $\mathrm{g/cm}^3$. Production runs were subsequently performed for 70 ns in the NVT ensemble.  In these simulations, the time step was 2 fs using bond length constraints through LINCS for all hydrogens bonded to heavy atoms.  Electrostatic calculations were performed using the Particle Mesh Ewald method with a 1.2  \AA{} cutoff in a rectangular box.  The van der Waals term was calculated using a potential switch between 1.0 and 1.2  nm.  We employed the modified CHARMM force field developed by Abel and coworkers\cite{Abel2004} for AOT surfactant and isooctane solvent molecules.  A list of $n+1$ systems studied is provided in ele \ref{compositions} with accompanying $R_g$ and $As$.  Images were rendered by VMD\cite{HUMP96}.  GROMACS 5.1\cite{Berendsen1995,Pall2015,abraham2015gromacs} was used to perform system preparation, minimization, and molecular dynamics simulation.  Structural and statistical analysis was performed using R\cite{Rpackage2013,ihak:gent:1996,rgl}.

Production simulations of ($n$+1)-mer systems were 70 ns length, employing a sampling interval of 10 ps. Production simulations of $n$-mer reference systems were 70 ns in length and employed a sampling interval of 4 ps. Only the last 50 ns of both of these simulations were used for analysis with ERmod. The structure of the AOT monomer reference system was performed for 70 ns with a sampling of 0.4 ps and only the last 5 ns of this simulation were used for test insertions.

Following simulation we found that dRM aggregates formed irregular, non-spherical shapes. As is often done in studies of micelle structure, we quantified the  deviation in dRM shapes from a perfect sphere by calculating the asphericity ($A_s$) of dRMs in the last 50 ns of all simulated systems. For a RM of total mass, $M$, the principal moments of inertia, $I_1~ > ~I_2~ >I_3$, are related to the semiaxes $a, b, c$  defined as
 \begin{eqnarray}
   I_1 &= \frac{1}{5} M (a^2 + b^2), &a^2 =\frac{5}{2M}(I_1+I_2-I_3)\\
   I_2 &= \frac{1}{5}M  (a^2 + c^2), &  b^2 = \frac{5}{2M}(I_1-I_2+I_3)\\
   I_3 &=\frac{1}{5} M (b^2 + c^2) ,&c^2 = \frac{5}{2M}(-I_1+I_2+I_3)
 \end{eqnarray}
 For the semiaxes $a>b>c$ of an ellipsoid, $A_s$ is defined 
    \begin{equation}
      A_s = \frac{\lambda_{z}^2 - \frac{1}{2} (\lambda _{x}^2 + \lambda_y ^2 )}{\lambda _{z^2}}, ~\lambda_x ^2 \leq  \lambda _y ^2 \leq  \lambda  _z ^2 ,
    \end{equation}
where $ A_s = 0$ for a sphere.
 The radius of gyration $R_g$ is defined as
 \begin{equation}
   R_g^2= \frac{\sum_{i}m_i (r_i - r_{com})^2 }{M} = \frac{a^2+b^2+c^2}{5},
 \end{equation}
 where $m_i$ is the mass of atom $i$ at distance $r_i$ from the RM's center of mass $r_{com}$. 
 The average radius of the dRM is $(abc)^{\frac{1}{3}}$ corresponding to the radius of a sphere with the same volume.

\section*{Results}
\label{sec:orgheadline7}

We characterized AOT aggregates of $n$-components in isooctane solvent, forming ``dry'' reverse micelles (dRM) of size $n$.  The radius of gyration ($R_g$), asphericity ($A_s$), and mass distribution functions of these dRMS provide insight into their internal structure.  Table \ref{compositions} lists the compositions of each simulated system, defined by the number of AOT molecules and the overall number of isooctane molecules used in the free energy evaluation.   As expected, the $R_g$ increases as $n$ increases, and $A_s$ approaches zero at large $n$.

\begin{figure}[htb]
\centering
\includegraphics[width=0.45\linewidth]{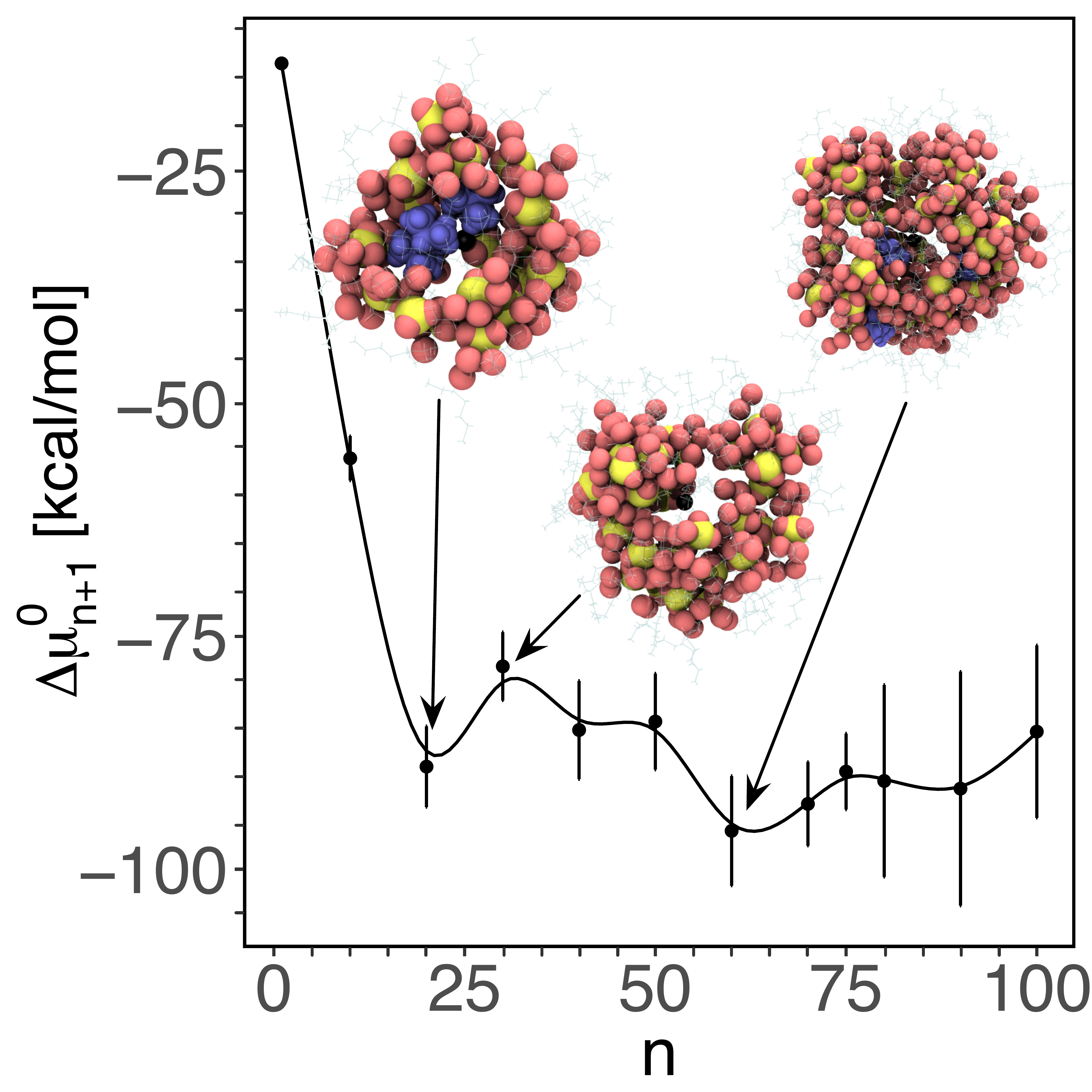}
\caption{\label{mu0}
The chemical potential, representing the change in free energy at standard state for a surfactant molecule to be added to a preexisting AOT $n$-mer.  Representative structures are shown for $n$=$20$,  $n$=$30$ and $n$=$60$.  The black sphere is the center of mass of the aggregate. Oxygen and sulfur atoms of AOT head groups are represented by a CPK model while the aliphatic surfactant tails are represented by lines.  The blue molecule in $n$=$60$ is an isooctane molecule confined in the core of the aggregate.}
\end{figure}

A principal result of this study is the determination of the chemical potential at standard state (infinitely dilute solution) for the process of adding one surfactant, $A$, to an existing micelle aggregate: $R_n$, $R_{n} + A \overset{\Delta \mu_{n+1}^{0}}\rightleftharpoons R_{n+1}$.  $\Delta \mu_{n+1}^{0}$, derived from Eq. (\ref{eq:1}), is shown in Fig. \ref{mu0} as a function of aggregate number $n$.  The changes in free energy were computed approximately using the ER method and were interpolated in order to determine values of the change in chemical potential for intermediate values of $n$ (See Methods). 

The overall profile of $\Delta \mu_{n+1}^{0}$ shows a rapid decrease with increasing aggregate size for small $n$, followed by a minimum near $n$=20 and subsequent increase to intermediate values near $n$=30.   As $n$ increases further, a broad second minimum is identified followed by a plateau for larger $n$.   Insight into the non-monotonic behavior of the free energy is provided by the representative structures associated with the two minima and intermediate maximum. Clearly, the dry micelle aggregates possess a structure that is not well captured by the radius of gyration and asphericity alone.

\begin{figure}[htb]
\centering
\includegraphics[width=1.0\linewidth]{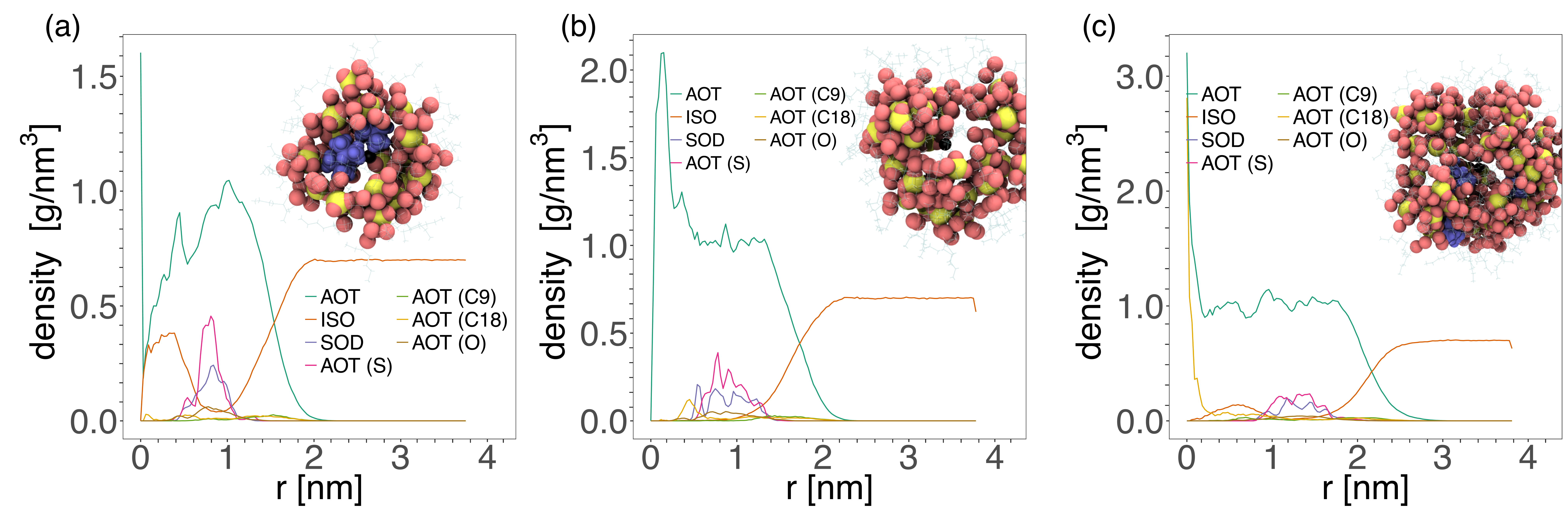}
\caption{\label{dens}
The mass density of many surfactant components are depicted as a function of distance from the center of mass of the dRM for \(n\) = (a) 20, (b) 30, and (c) 60. Oxygen and sulfur atoms of AOT head groups are red and yellow, respectively, and aliphatic surfactant tails are transparent gray.  The isooctane molecules occupying the dRM core are shown in blue.}
\end{figure}

Further insight into the structure of aggregates of varying size is provided by analysis of the internal mass density distribution.
Fig. \ref{dens} depicts the mass density of several atom groups as a function of distance from the center of mass of the dRM. 
The mass distributions suggest that the abrupt change in free energy between $n$=20 and $n$=30 results from the aggregation of head groups near the center-of-mass of the dry micelle.

In the $n$=20 dRM AOT head groups colocalize near the center-of-mass leading to a peak in the sulfur atom mass distribution near 8\  \AA, and oxygen and sodium peaks also appear near 8\  \AA.  As the representative inset structure shows, the head groups form a folded ``taco shell'' conformation rather than a spherical aggregate. This allows for the formation of a compact micelle structure with head groups packed near the micelle center while minimizing head group repulsion.  A few isooctane molecules were found to be filling the core of the dRM taco shell rather near the AOT tails. Such penetration is possible as a result of the small difference in surface tension between organic solvent, 25-35 mN/m\cite{schick1987nonionic,hollamby_effect_2008}, and AOT,  30 mN/m\cite{nave2000so}. (This can be compared to the value for water,  ~70 mN/m.)  
Given the small difference in surface tension, structural fluctuations that allow the dRM center to have both AOT tail and isooctane solvent are possible. This minimum at $n$=20 is a bit larger than that predicted by Eskici and Axelsen, $n$=13.6, found by extrapolation of a relation of  numbers of water, surfactant, and salt in spherical RMs from simulations at water loading 7.5 in similarly dilute conditions. Eskici and Axelsen's extrapolation might otherwise hold if we did not observe such complex aggregate shapes in these dry conditions, which allow for a higher degree of AOT aggregation.

The $n$=30 dRM is less stable than other slightly larger or smaller dRMs, as $\Delta \mu_{30+1}^{0}$ is a local maximum in Fig. \ref{dens}b. Head group repulsion leads to a more complex dRM structure.  While the overall asphericity of the aggregate is small (0.21), the internal structure associated with the arrangement of surfactant head groups creates two joined tori.  This internal structure reflects a mass distribution with a strong peak in AOT tail group density near 5 \AA, with a wider sulfur atom distribution between 6 \AA \ to 12 \AA.  This suggests that the shift in the position of the AOT sulfur atoms make an important contribution to the free energy surface maximum. It seems that the identity of dRM core molecules (isooctane or AOT tails) has little effect on the interaction energy in comparison to the electrostatic energy of surfactant head group interactions.

As the micelle size increases further to $n$=60, at which point we observe the global minimum in chemical potential, a second morphological transition occurs in Fig. \ref{mu0}.  A spherical shell of AOT head groups form, reflected by a peak in the sulfur atom mass density between 14 \AA  \ and 17 \AA in Fig. \ref{dens}c. There is a substantial density of surfactant tail groups and isooctane molecules in the center of the micelle, suggesting that the micelle core is composed of non-polar molecules in contrast to the structure of a ``wet'' reverse micelle.   

This arrangement of surfactant head groups, distributed over a spherical shell containing surfactant tail groups as well as non-polar solvent, is reflected in a decrease in $\Delta \mu_{n+1}^0$. After $n$=60, $\Delta \mu_{n+1}^0$ is observed to plateau at a constant value within the error bars of our computed values.

The observed penetration of non-polar solvent molecules has been proposed for wet reverse micelles in benzene solvent based on NMR experiments\cite{martin1981carbon}.   However, the inclusion of non-polar solvent in the core of dry AOT RMs has not been previously observed or proposed. 
Moreover, the structural fluctuation of solvent molecules after the formation of the micelle is consistent with the unfavorable entropy and enthalpy differences observed in experiment. In normal micelles, after micelle formation fluctuations in surfactant molecules decrease while fluctuations in water molecules increase, leading to an increase in entropy during micellization. However, in dRM formation micellization causes surfactant molecules to \emph{lose} translational entropy, leading to a decrease in total entropy. This provides an explanation for the negative entropy change associated with RM micellization.
It appears that the structure of larger dRMs balances the favorable aggregation of surfactant while minimizing the penalty of electrostatic repulsion among charged anionic head groups.  Upon reaching a critical size, the dRM is able to form a spherical structure characterized by a shell of head groups with non-polar surfactant tails facing outward from the dRM center, encapsulating a non-polar solvent core.  This addresses the long-standing question regarding how anionic AOT surfactant molecules can form reverse micelle structures in non-polar solvent in the absence of a cosurfactant or water.

This work thus far has only considered AOT dRMs in near infinite dilute condition, such that there is no contribution to the chemical activity from inter-micellar interactions.  Considering the activity coefficients for AOT reverse micelles, we expect that molecular charging will require higher energy cost in non-polar as opposed to water solvent. In particular, the anionic AOT molecules strongly mediate such charging in non-polar solvent\cite{hsu2005charge,briscoe2002direct}, and we expect that electrostatic interaction among RMs strongly affects their size distribution.
Similar electrostatic interactions have been examined in various experiments and explained in terms of Debye-H{\"u}ckel theory and its screened potential form\cite{dufresne2005reverse,pinero2004electrostatic,roberts2008electrostatic,sainis2008electrostatic}. 
Moreover, the charge fluctuation theory by Eicke et al., which assumes no electrostatic interaction was found to agree with wet RM experimental data only at relatively higher water concentrations, failing as water concentrations approached dry conditions. Those authors assigned the observed deviation to inter-micellar electrostatic interactions in dry conditions\cite{eicke1989conductivity}, implying that dRMs are charged during collisions at equilibrium. These past observations in combination with our observation of dRM size and shape (controlled by intra-aggregate electrostatic repulsions) suggest that Debye-H{\"u}ckel theory can provide an accurate description of the activity. As such, we  assume
\begin{equation}
 \log \left( \frac{\gamma _{n} \gamma _{1} }{\gamma _{n+1} } \right)  \propto \alpha n, \label{eq:ratio}
\end{equation}
implying a simple relation between representative charge and aggregate size, $\log{\gamma_{n}} \propto \alpha^{\prime} q^2 = \alpha  n^2$ as previously used for normal micelles\cite{Yoshii2006b}. We note that there is no molecularly-detailed information available for the inter-micellar charges felt by dRMs as a function of size. As such, we assume that the charge and size are proportional to each other for simplicity.

With this assumption, $\alpha$ determines how much the charge of dRMs increase as a function of size, which may depend on conditions such as the identity of non-polar solvents, the ionic strength resulting from head group charges and counter ion type, and temperature.  This scaling is expected to be valid in dilute neutral electrolyte solution for concentration on the order of  \(0.1\) M.  To investigate the effect of varying activity coefficient, we explored a range of $\alpha$, $-\infty <\alpha <0$ ($0<\frac{\gamma _{n} \gamma _{1} }{\gamma _{n+1} }<1$). In this range $\alpha$=$0$ represents a ``phase separated'' state, where the largest possible dRM is formed from all surfactants in the solution. $\alpha$=$-\infty$ represents a ``salt-out'' state where only monomers are formed in solution. Somewhere within this range, we expect a value of $\alpha$ which produces a ``micellar'' solution, exhibiting a critical micelle concentration near our first local minimum in $\Delta \mu_{n+1}^{0}$.  Using this assumed relation allows us to probe the effect of interaggregate interactions for a given size distribution. 

We rewrite the size distribution given in Eq. (\ref{eq:1}) relative to an assumed monomer mole fraction, employ our assumed value for $\alpha$, and multiply smaller size relations iteratively transforming Eq. (\ref{eq:1}) into

\begin{equation}
  \log{X_{n}} = n \log{X_{1}} + \alpha \sum_{ i=1}^{n-1} i -  \frac{\sum_{i=2}^{n} \Delta \mu ^{0}_{i}   }{k_{B}T}. \label{eq:2}
\end{equation}

\begin{figure}[htb]
\centering
\includegraphics[width=0.45\linewidth]{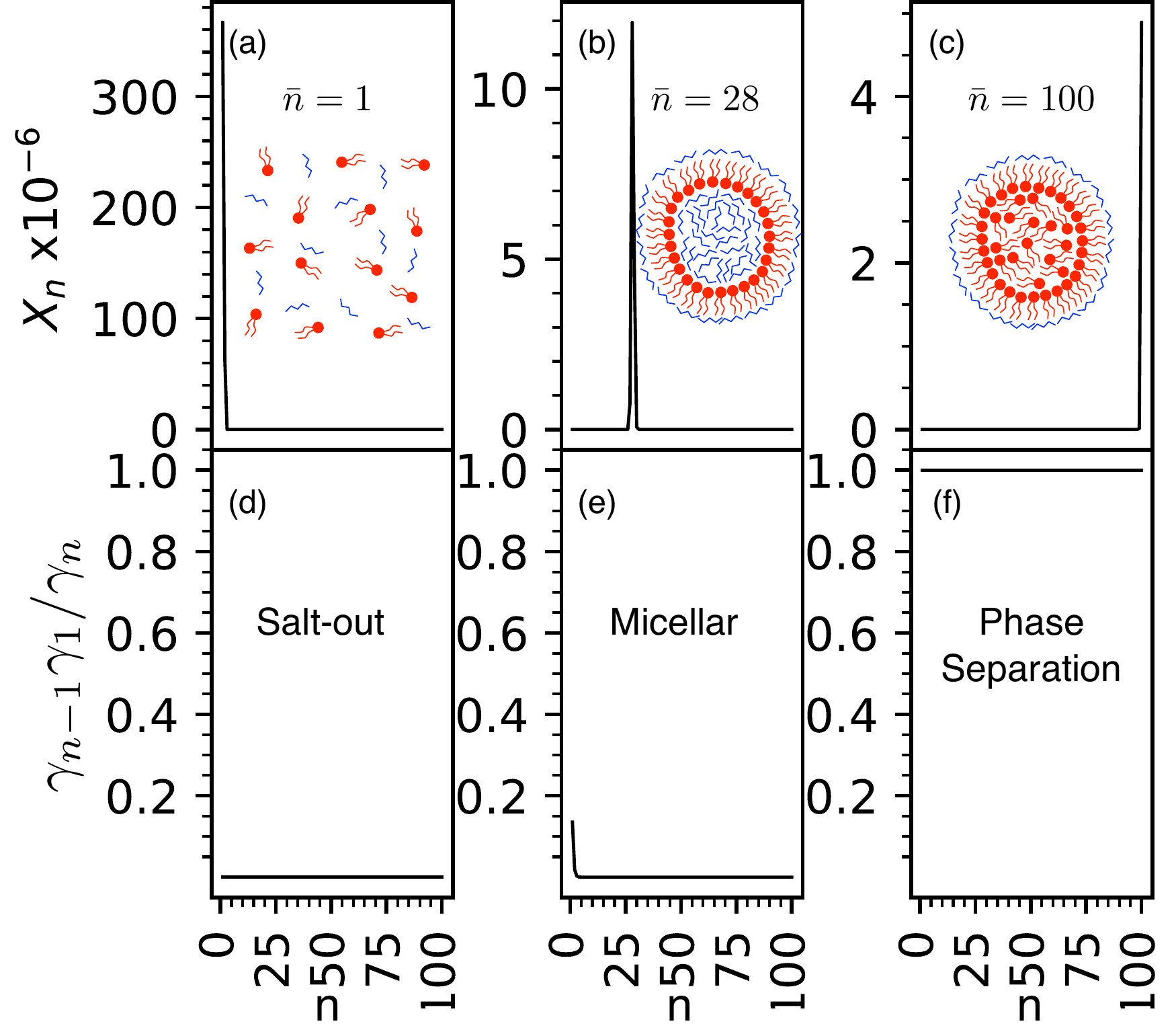}
\caption{\label{dists_gamma1}
The surfactant aggregate number distributions produced by (a) $\alpha$=0, (b) $\alpha$=-2, and (c) $\alpha$$\simeq$$-\infty$ corresponding to salt-out, micellar, and phase separated dRM solutions in systems containing up to 100 surfactant molecules.  Below, (d), (e), and (f) are the activity coefficient ratios as a function of \(n\) for the corresponding size distributions in (a), (b) and (c). In (c) $\bar{n}$ will be the total of all surfactant molecules in the system, forming a single aggregate. Cartoons depict surfactant (red) and non-polar solvent (blue).}
\end{figure}

This equation displays how the free energy surface and $\alpha$ determine the mole fraction of aggreagtes of size $n$. Using this relation, we demonstrate the values of $\alpha$ that produce the aforementioned ``salt-out'', ``micellar'', and ``phase separated'' distributions of $X_n$ Fig. \ref{dists_gamma1}. The ``salt-out'' and ``phase separated'' conditions correspond to $\alpha$$\simeq$$-\infty$ and $\alpha$=0. We discovered that a value of $\alpha$=-2 produces an apparent critical micelle concentration at $\bar{n}$=28. The actual solution of AOT surfactant in isooctane solvent must balance electrostatic repulsions and entropic penaltiesf resulting from the aggregate concentration in the equilibrium state\cite{hsu2005charge}. This implies that larger aggregates have strong electrostatic repulsion while monomer and smaller aggregates are stabilized by entropy. These results suggest that interaction between dRMs are sensitive to AOT surfactant concentration. 

\begin{figure}[htb]
\centering
\includegraphics[width=0.45\linewidth]{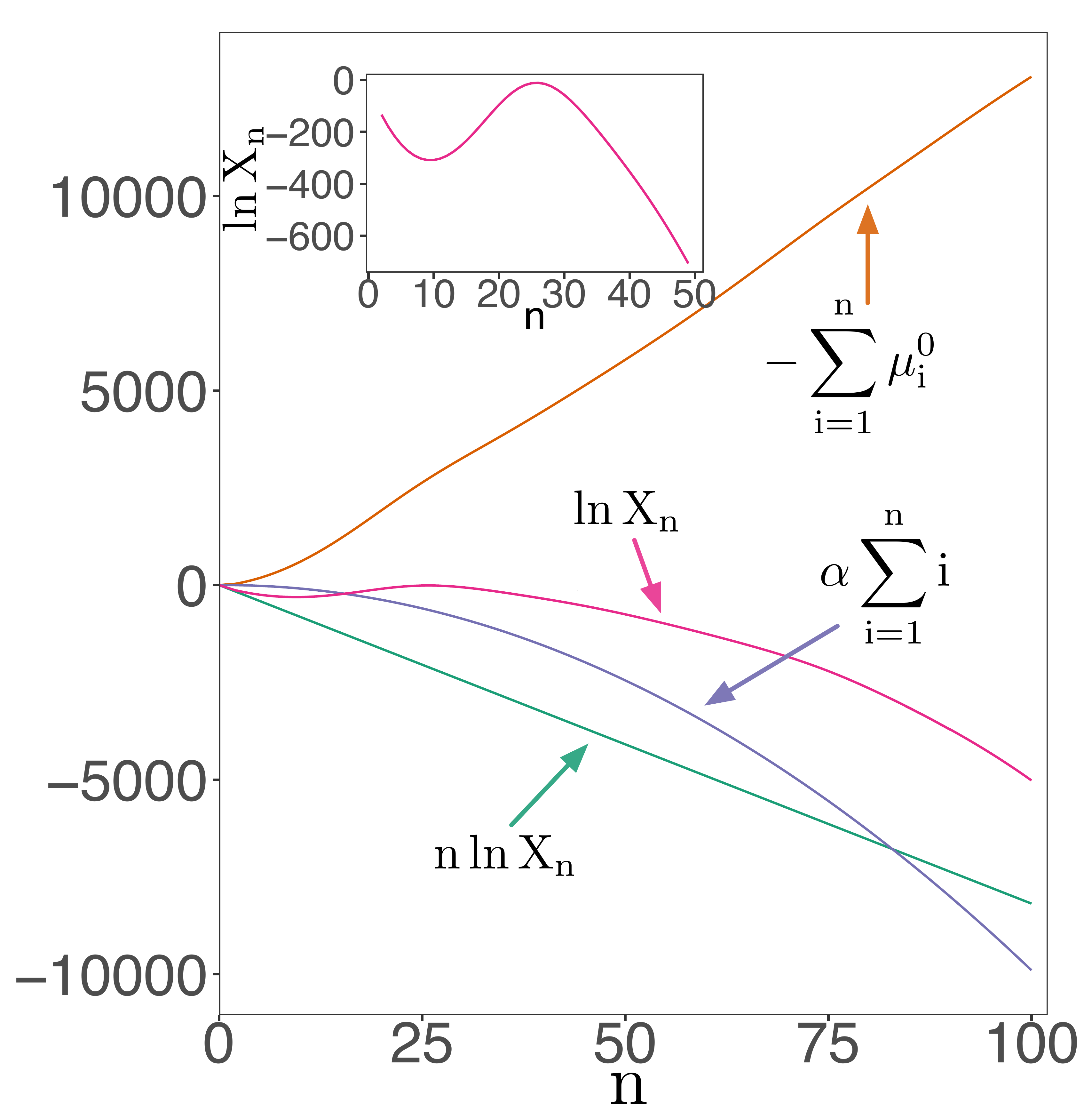}
\caption{\label{cumXn}
Individual terms contributing to Eq. (\ref{eq:2}) as a function of aggregate size.  The inset figure shows the expansion of  the logarithm of the aggregate mole fraction. $\alpha$, is set to $-2$. The $X_1$ value is set such that $c$=$c^m$.
}
\end{figure}

To better appreciate the aggregate size distribution depicted in Fig. \ref{dists_gamma1}(b), the balance of each term in Eq. (\ref{eq:2}) is shown in Fig. \ref{cumXn}. $\alpha$ was set to -2 and the total mole fraction was conserved at the CMC for the $X_{1}$ term. These separate contributions show that the finite aggregation at $n$=28 occurs due to the convex nature of the cumulative sum of free energy from the local minimum at $n$=20 through the local maximum at $n$=30. The narrow distribution is a direct reflection of the height of the barrier separating the local minima. The height of the barrier is on the order of several tens of kcal/mol resulting in a relatively monodisperse aggregate size distribution characteristic of a critical micelle concentration.

Rather than impose a constant $\alpha$ for any total concentration of AOT, as in Eq. \ref{eq:ratio}, it is more physically meaningful to introduce penalties to the size distribution of dRMs when surfactant concentrations exceed the critical micelle concentration, $c^{m}$. A small change in the mean aggregate size after the CMC was reached has been observed in experiments\cite{smith2016effect}.  This implies that near the CMC the activity coefficient abruptly increases from near zero as a function of concentration, and may subsequently be considered essentially constant. While normal micelle formation is dominated by the free energy difference between dispersed aggregated states and micelle states, it seems that interactions among dry micelles are a driving force that determines the equilibrium state size distribution.  The importance of micelle-micelle interactions for the observed phase transition near the CMC has been previously suggested\cite{de_solution_1995,mathew_concept_1988,grest_dynamic_1986,huang_attractive_1984}.  However, to our knowledge, no previous theory has captured the difference in free energy associated with intermicelle interaction.

As such, we employ an interpolation of $\alpha$ between values representing the salt-in state ($\alpha_{salt-in}$) and micelle state ($\alpha_{micellar}$) using a sigmoidal function to reproduce the sudden occurrence of micelles near the CMC as a function of the surfactant concentration ($c$=$\sum_{n=1}nX_n$) expressed as
\begin{equation}
\alpha(c) =  A + \frac{( \alpha_{salt-in} - \alpha_{micellar})}{ [1 + \exp{(-A (c - c^{m}}))]^{(B^{-1})}}
\label{eq:slope}
\end{equation} 
where $A$ and $B$ scale the sigmoid form, set to $A$=$10^6$, and $B$=35. We set  $\alpha_{salt-in}$=-30 and $\alpha_{micellar}$=-2.

\begin{figure}[htb]
\centering
\includegraphics[width=0.45\linewidth]{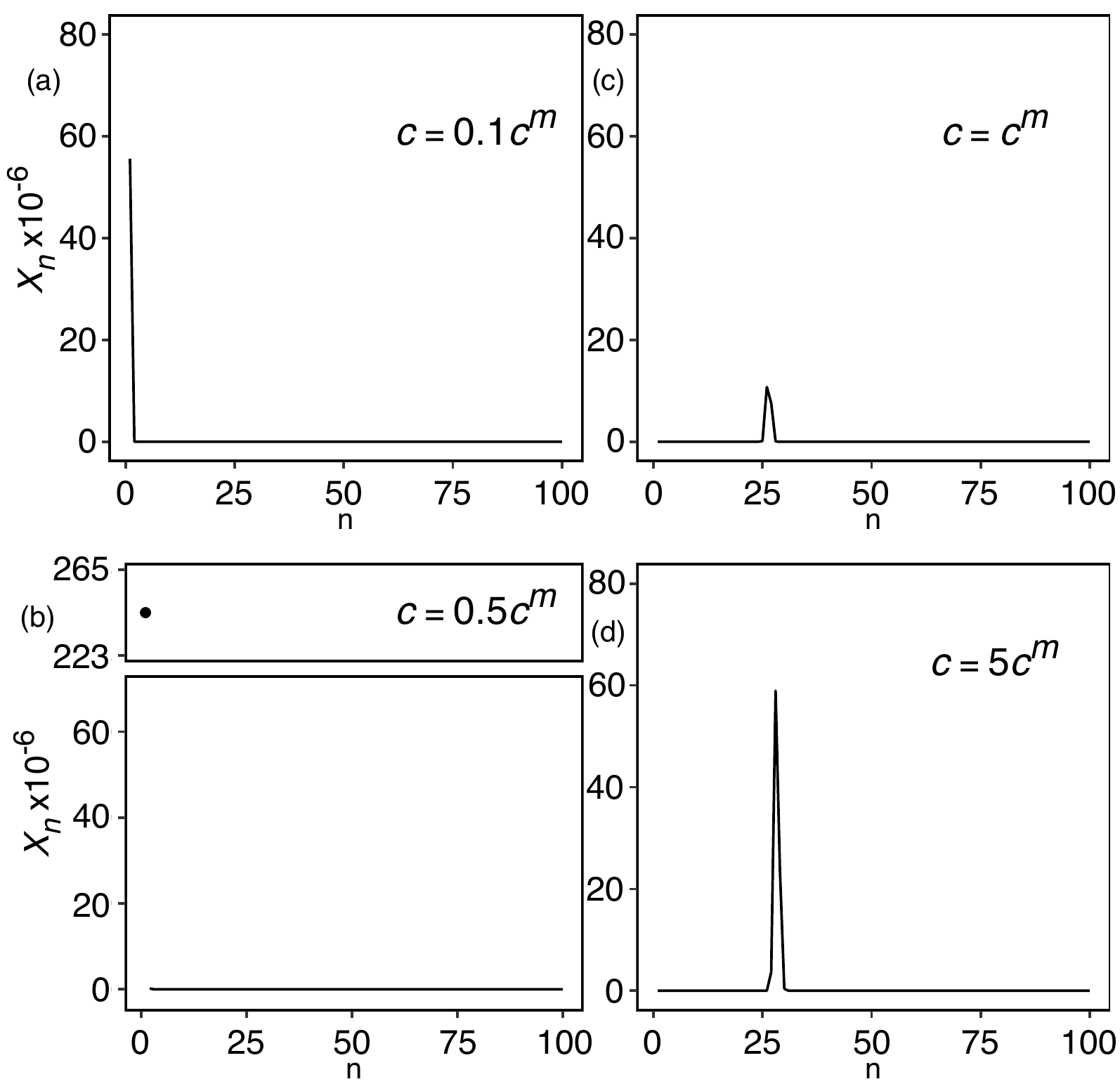}
\caption{\label{dists}
The micelle size distribution is shown as a function of total concentration of AOT molecules for four values of the critical micelle concentration ($c^{m}$), defined in Eq. (\ref{eq:slope}).  The parameter $\alpha$, which defines the size scaling of the ratio of activity coefficients (Eq. \ref{eq:ratio}), was fixed to form micelles at a $c^{m}$=0.0049 M.}
\end{figure}

Using this interpolated function $\alpha(c)$, we test how various values for the ratio of CMC ($c^{m}$) to sufactant concentration ($c$) control the size distribution. Fig. \ref{dists} shows the micelle size distribution for surfactant concentration varying from $c$=$0.1c^{m}$ to $c$=$5c^{m}$.  The resulting micelle size distributions are narrow with width $\Delta n$=2, which is consistent with experimental results for an AOT/alkane system\cite{Smith2013}.  

For solutions of normal micelles, it is observed that as the concentration increases near the CMC the size distribution shows a gradual increase from monomer to normal micelle.\cite{diamant2016free} For the dry reverse micelle, we observe that surfactant AOT molecules make a sudden transition from monomer to micelle aggregate near the CMC.  Such an abrupt transition from the monomer regime to the micelle state has been anticipated but never observed in normal micelle solutions. As such, this represents a striking difference between the thermodynamics of normal micelle formation and the formation of dry RMs.  
The observed difference results from the magnitude of variation in $\Delta {\mu}^{0}_{n+1}$ over several kcal/mol in normal micelles and several tenths of kcal/mol in the case of dry RMs.
Variations in enthalpy over tenths of a kcal/mol have been observed in an experimental study of AOT micellization in alkane solvent \cite{mukherjee1993thermodynamics}.  As the enthalpy change in surfactant AOT micellization depends on the choice of solvent, we expect a corresponding dependence of the RM size distribution on the choice of solvent.

\begin{figure}[htb]
\centering
\includegraphics[width=0.45\linewidth]{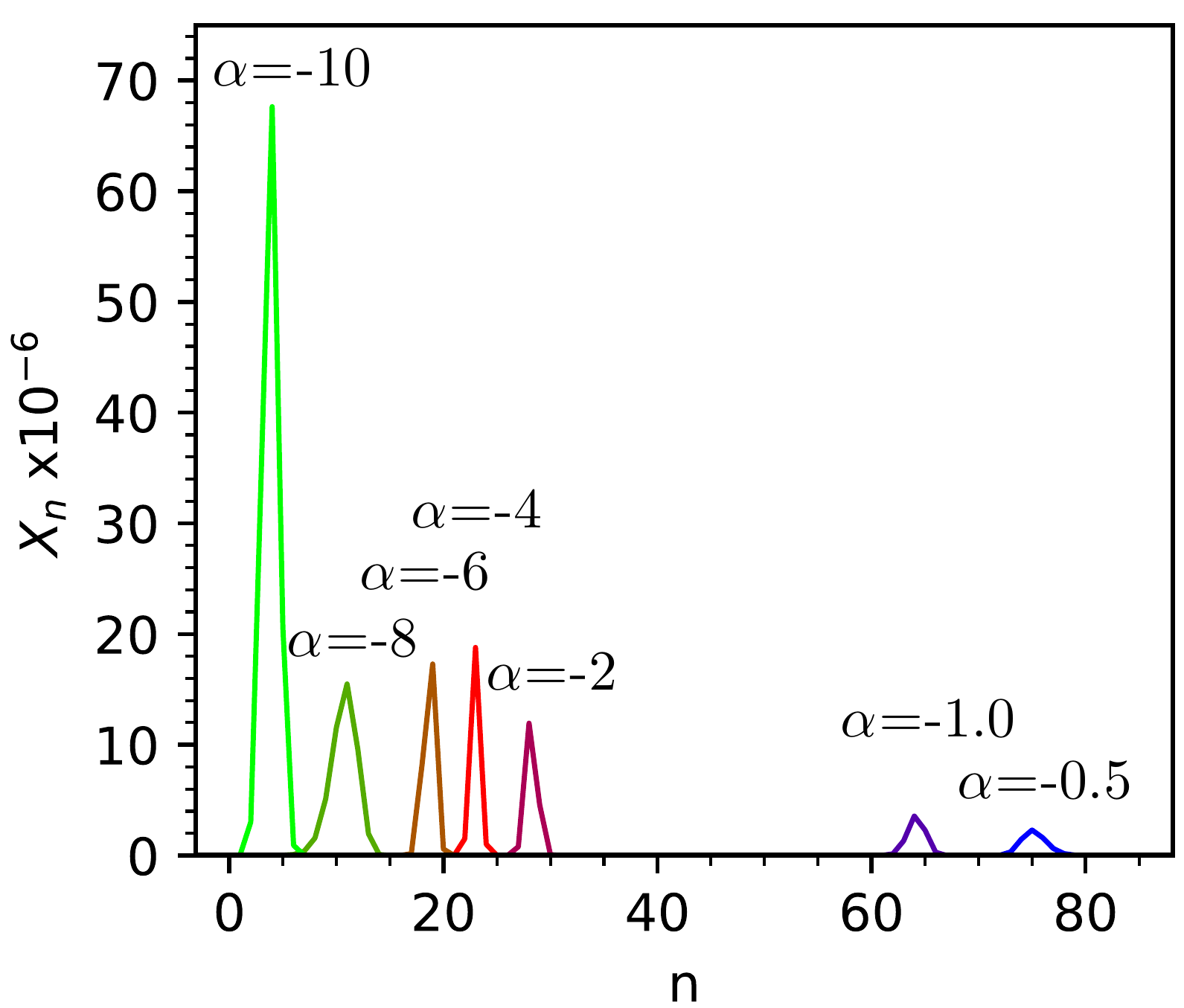}
\caption{\label{slopes}
The aggregate size distribution for maximum value of $\alpha(n)$ defined in Eq. \ref{eq:slope}.  Values span a range demonstrating the regime between ``salt-out'' conditions and ``phase separated'' conditions within which we find micellar conditions. The total concentration is fixed to be the CMC.}

\end{figure}

 Note that the observed transition is invariant to the specific nature of the function used to model $\alpha$.  Fig. \ref{slopes} shows the continuous change in the micelle size distribution as a function of $\alpha$.  The mean size of the dry micelle is observed to depend on the value of $\alpha$, while the point at which AOT molecules form aggregates from monomers is relatively insensitive to $\alpha$. This suggests that the sigmoidal function employed in this work to interpolate values of the activity coefficient up to the CMC, given by Eq. (\ref{eq:slope}), does not influence the specific regime in which the first dry RM is observed.

In addition, the mean aggregate size, $\bar{n}$, for the distributions gradually increases from 1 in salt-out conditions through $\bar{n}$=30 by varying $\alpha$ from $-\infty$ to -2.  This behavior is in agreement with the experimentally reported values of surfactant AOT ranging from  $\bar{n}$=30 in {\it n}-octane, to 37 in {\it n}-decane, to $\bar{n}$=44 in {\it n}-dodecane \cite{Frank1969,eicke1980surfactants}. This implies that if the activity coefficient gradually increases as a function of AOT concentration, the system will display no CMC in this region. Consider the gradual increase observed in $\bar{n}$, which shows a sharp transition for 30 to 65 (Fig. \ref{slopes}). Moreover, regions of higher AOT concentration may stand beyond the dilute solution limit.  These specific aggregate sizes correspond to local maximum values in $\Delta \mu_{n+1}^0$. We must note, however, that our results for the mean free energy are accompanied by large error bars.  As such, the transition from 30 to 65 may be an artifact.  Assuming that the free energy surface is constant after $n$=50 in $\Delta \mu_{n+1}^0$, our results imply that the surfactant AOT molecules become insoluble near $n$=100, in agreement with the liquid crystal state of the ternary phase diagram along the AOT/isooctane line\cite{tamamushi1980formation}.

\section*{Discussion}
\label{sec:orgheadline8}

In this study, we have obtained the size distribution of dry surfactant AOT aggregates in non-polar solvent. The free energy of the isolated aggregate formation was evaluated through free energy simulation using the energy representation method.  The obtained free energy surface demonstrates that small surfactant aggregates can form, resulting in a dense surfactant head group region.  In surfactant aggregates, the AOT heads groups disperse from the center of mass of the aggregate and create a core region that can accommodate isooctane solvent molecules. The spontaneous formation of a pore-like ``taco shell'' structure allows for minimization of head group repulsion and the formation of stable dry RM aggregates.  In modeling the interaction of dry RM aggregates, we employed the Debye-H{\"u}ckel theory scaling relation for variation in the activity coefficient with increasing aggregate number in dilute solution.  It was observed that depending on the degree of interaction, the dry RM size distribution displayed a variety of aggregate forms including soluble dry micelles and insoluble aggregates.

Previous theoretical work has doubted, the existence of a CMC for dry RMs\cite{ruckenstein1980aggregation,Wootton2008}. We believe this was a direct result of an inability to accurately describe the free energy of solution of dry RM aggregates assuming ideal solution conditions. 
Recently, Kislenko and Razumov\cite{kislenko2017molecular} investigated dry RM formation in an AOT/hexane system using thermodynamic integration and all-atom simulations.  Their free energy surface showed features similar to the dependence reported here.  However, their final  size distribution contained dRMs of finite size, but these were found to be roughly twenty orders of magnitude smaller than that of the monomer, and much lower than that of the largest aggregates which had formed due to ideal solution conditions.  This demonstrates that one must account for interactions between surfactant aggregates.
In this work, we obtained the formation free energy of aggregates through free energy evaluation using the energy representation method. The mean aggregate size was found to increase as a function of total surfactant concentration in a way that depends on variation in the activity coefficient.

If the activity coefficient displays a weak increase as a function of concentration, the mean aggregate size displays a gradual increase from $n$=1 to 30.  This dependence is determined by the variation in the chemical potential $\Delta \mu^{0}_{n+1}$ as a function of dry RM size.  For aggregates larger than 30, the mean aggregate size was observed to undergo an abrupt transition to much larger aggregate sizes. This dependence results from the variation in the monomer activity coefficient with an increase from 0 (an extreme ``salt-out'' condition) to 1 (an ideal solution condition). 

From this variation, we conclude that the interaction of surfactant aggregates in non-polar solvent plays a major role in dry RM formation.  This behavior is quite distinct from normal micelle formation, approximately expressed using ideal solution conditions.  Moreover, our results suggest why calorimetry experiments\cite{TANAKA1992} have failed to observe the existence of a CMC for dry AOT surfactant RM formation.

The computed free energy surface suggests that there will be only a minor change in aggregate size distribution as a result of a modest increase in temperature of the RM phase. Note that at the threshold temperature separating the RM and liquid crystal states intermicelle interaction is found to be critical. This implies that a change in temperature cannot trigger a shift in the equilibrium state between a distribution dominated by surfactant monomers and one defined by aggregates.  As such, we do not observe a maximum in the heat capacity as a function of surfactant concentration.  

Our results also suggest a reason for the observed deviation in mean aggregate size in experiment.  In larger aggregates \(n>20\), we observed the penetration of isooctane molecules in the center of the surfactant aggregate, with the solvent ``core'' surrounded by aliphatic tail groups of the AOT surfactant. In experimental estimates of the mean size of surfactant aggregates, it is typically assumed that the aggregate is solely composed of AOT surfactant molecules. This assumption may lead to an overestimate in the aggregation number.  In the case of dry RM formation by AOT surfactant in isooctane solvent, we estimate that this overestimate of the aggregation number can be as large as 10. In addition, many experimental studies assume a spherical surfactant aggregate.  However, we observe that due to the formation of the solvent core large aggregates may deviate from a spherical shape. Moreover, the penetration of non-polar solvent in dRMs may cause a decrease in entropy upon micellization.

We compute the chemical potential in the infinite dilution limit, $\Delta \mu^0_{n+1}$. We have developed a theory which separates ``ideal'' (intra-micellar) and ``non-ideal'' (inter-micellar) contributions to the free energy as dependent on surfactant concentrations which determine the dRM size distribution. The following is a brief discussion of how these contributions inform the total free energy.

Consider the reaction $R_n  + A \rightleftharpoons R_{n+1}$.
The equation $\mu_n = \mu^0_n + kT  \ln a_n = \mu^0_n + kT  \ln (\gamma_n X_n)$ demonstrates that the free energy consists of two terms, an ideal contribution, $\mu^{ideal}_n = \mu^0_n + kT \ln X_n$, and non-ideal contribution, $\mu^{non-ideal} = kT \ln a_n$.
The non-ideal contribution can be written
$kT \ln \gamma_n = (H_n - H_n^{ideal}) - T (S_n - S_n^{ideal})$,
where $H_n$ and $S_n$ are the enthalpy and entropy of a micelle of size $n$, respectively.
If the size and morphology of a micelle of size $n$+1 do not show significant change between the standard state and actual solution state, we expect that the entropic contribution to the non-ideal component of $\ln (\gamma_n)$ will be modest.
In contrast, the standard state enthalpic contributions from intermicelle interactions involving aggregates of size $n$ and $n+1$ are not expected to be significant compared to the contribution from micelle-solvent interaction.   However, at higher concentration the binding reaction is completed in the presence of other micelles.
As such, the major contribution of $H_n - H_n^{ideal}$ to $\gamma_n$  is due to the interaction among micelles. (The interaction between micelles of size $n$ and solvent is already included in $H_n^{ideal}$.) No other difference exists between the chemical potential of the standard state, $\mu_n^0$, and the actual state, $\mu_n$.  This reasoning demonstrates that the change in activity coefficient includes contributions from intermicelle interaction as a function of AOT concentration.

We make one further observation.  We have employed the chemical species model for the derivation of our thermodynamic relations.  An assumption of the model is that 
the activity coefficient $\gamma_n$ is a function of the distribution of sizes of other micelles present in the solution or that
\begin{equation}
  \ln \gamma_n = f([RM_1], [RM_2], ..., [RM_{n-1}], [RM_{n+1}], ..., [RM_{N_{AOT}}] )
\end{equation}
rather than
$\ln \gamma_n = f([AOT])$ which includes solute-solute interactions.
On the other hand, as the total concentration of surfactant $[AOT]$ increases, it will change the distribution of micelle sizes and the values of $[RM_1], [RM_2], ..., [RM_{n-1}], [RM_{n+1}], ..., [RM_{N_{AOT}}]$, leading to a change in the activity for each micelle size.
This suggests that $\gamma_n$ should vary explicitly as a function of surfactant concentration for wider regions of total $[AOT]$. Such strong intermicelle interaction is observed in experiments\cite{fletcher1987kinetics,mejuto2014effects} related to mass transport and percolation in wet RMs.

Finally, the current treatment of activity in Debye-H{\"u}ckel theory is the simplest one. Although variation of the parameter covers the region of experimental mean aggregate size from 30 to 60, a more refined model is necessary to understand how a specific experimental condition affects the mean aggregate size. The development of a more detailed interaction model for RM formation is a future goal for the field.

In summary,  our results suggest that in non-polar solvent, molecular interactions between surfactant aggregates and non-polar solvent molecules play an important role in establishing equilibrium for dry RM formation due to the fact that surfactant molecules demonstrate a weak ability to associate in non-polar solvent.   Addition of water molecules to a dry RM system leads to the formation of a wider variety of stable RMs.  The extension of the formalism developed in this work to the case of ``wet'' RMs, such as those found in a ternary AOT/isooctane/water system, should allow for a similar characterization of the equilibrium RM size distribution.

\section*{Acknowledgments}
\label{sec:orgheadline9}

We gratefully acknowledge the National Science Foundation (CHE-1362524) for the generous support of our research. G.A.P. thanks the National Science Foundation Graduate Research Fellowship Program (DGE-1247312). JES thanks the Japan Society for the Promotion of Science (JSPS) for the support provided by an Invitation Fellowship (L13523) and BRIDGE Fellowship (BR160401) hosted at Nagoya University.  

\bibliography{dry}

\begin{thebibliography}{87}%
\makeatletter
\providecommand \@ifxundefined [1]{%
 \@ifx{#1\undefined}
}%
\providecommand \@ifnum [1]{%
 \ifnum #1\expandafter \@firstoftwo
 \else \expandafter \@secondoftwo
 \fi
}%
\providecommand \@ifx [1]{%
 \ifx #1\expandafter \@firstoftwo
 \else \expandafter \@secondoftwo
 \fi
}%
\providecommand \natexlab [1]{#1}%
\providecommand \enquote  [1]{``#1''}%
\providecommand \bibnamefont  [1]{#1}%
\providecommand \bibfnamefont [1]{#1}%
\providecommand \citenamefont [1]{#1}%
\providecommand \href@noop [0]{\@secondoftwo}%
\providecommand \href [0]{\begingroup \@sanitize@url \@href}%
\providecommand \@href[1]{\@@startlink{#1}\@@href}%
\providecommand \@@href[1]{\endgroup#1\@@endlink}%
\providecommand \@sanitize@url [0]{\catcode `\\12\catcode `\$12\catcode
  `\&12\catcode `\#12\catcode `\^12\catcode `\_12\catcode `\%12\relax}%
\providecommand \@@startlink[1]{}%
\providecommand \@@endlink[0]{}%
\providecommand \url  [0]{\begingroup\@sanitize@url \@url }%
\providecommand \@url [1]{\endgroup\@href {#1}{\urlprefix }}%
\providecommand \urlprefix  [0]{URL }%
\providecommand \Eprint [0]{\href }%
\providecommand \doibase [0]{http://dx.doi.org/}%
\providecommand \selectlanguage [0]{\@gobble}%
\providecommand \bibinfo  [0]{\@secondoftwo}%
\providecommand \bibfield  [0]{\@secondoftwo}%
\providecommand \translation [1]{[#1]}%
\providecommand \BibitemOpen [0]{}%
\providecommand \bibitemStop [0]{}%
\providecommand \bibitemNoStop [0]{.\EOS\space}%
\providecommand \EOS [0]{\spacefactor3000\relax}%
\providecommand \BibitemShut  [1]{\csname bibitem#1\endcsname}%
\let\auto@bib@innerbib\@empty
\bibitem [{\citenamefont {Zhang}\ \emph {et~al.}(2007)\citenamefont {Zhang},
  \citenamefont {Guo}, \citenamefont {Dong},\ and\ \citenamefont
  {Sun}}]{zhang_characterization_2007}%
  \BibitemOpen
  \bibfield  {author} {\bibinfo {author} {\bibfnamefont {D.-H.}\ \bibnamefont
  {Zhang}}, \bibinfo {author} {\bibfnamefont {Z.}~\bibnamefont {Guo}}, \bibinfo
  {author} {\bibfnamefont {X.-Y.}\ \bibnamefont {Dong}}, \ and\ \bibinfo
  {author} {\bibfnamefont {Y.}~\bibnamefont {Sun}},\ }\href {\doibase
  10.1021/bp060188b} {\bibfield  {journal} {\bibinfo  {journal} {Biotechnol
  Progress}\ }\textbf {\bibinfo {volume} {23}},\ \bibinfo {pages} {108}
  (\bibinfo {year} {2007})}\BibitemShut {NoStop}%
\bibitem [{\citenamefont {Hino}, \citenamefont {Kawashima},\ and\ \citenamefont
  {Shimabayashi}(2000)}]{hino_basic_2000}%
  \BibitemOpen
  \bibfield  {author} {\bibinfo {author} {\bibfnamefont {T.}~\bibnamefont
  {Hino}}, \bibinfo {author} {\bibfnamefont {Y.}~\bibnamefont {Kawashima}}, \
  and\ \bibinfo {author} {\bibfnamefont {S.}~\bibnamefont {Shimabayashi}},\
  }\href {\doibase 10.1016/S0169-409X(00)00098-3} {\bibfield  {journal}
  {\bibinfo  {journal} {Adv. Drug Deliv. Rev.}\ }\bibinfo {series} {Emulsions
  for Drug Delivery},\ \textbf {\bibinfo {volume} {45}},\ \bibinfo {pages} {27}
  (\bibinfo {year} {2000})}\BibitemShut {NoStop}%
\bibitem [{\citenamefont {Okochi}\ and\ \citenamefont
  {Nakano}(2000)}]{okochi_preparation_2000}%
  \BibitemOpen
  \bibfield  {author} {\bibinfo {author} {\bibfnamefont {H.}~\bibnamefont
  {Okochi}}\ and\ \bibinfo {author} {\bibfnamefont {M.}~\bibnamefont
  {Nakano}},\ }\href {\doibase 10.1016/S0169-409X(00)00097-1} {\bibfield
  {journal} {\bibinfo  {journal} {Adv. Drug Deliv. Rev.}\ }\bibinfo {series}
  {Emulsions for Drug Delivery},\ \textbf {\bibinfo {volume} {45}},\ \bibinfo
  {pages} {5} (\bibinfo {year} {2000})}\BibitemShut {NoStop}%
\bibitem [{\citenamefont {Higashi}\ and\ \citenamefont
  {Setoguchi}(2000)}]{higashi_hepatic_2000}%
  \BibitemOpen
  \bibfield  {author} {\bibinfo {author} {\bibfnamefont {S.}~\bibnamefont
  {Higashi}}\ and\ \bibinfo {author} {\bibfnamefont {T.}~\bibnamefont
  {Setoguchi}},\ }\href {\doibase 10.1016/S0169-409X(00)00100-9} {\bibfield
  {journal} {\bibinfo  {journal} {Adv. Drug Deliv. Rev.}\ }\bibinfo {series}
  {Emulsions for Drug Delivery},\ \textbf {\bibinfo {volume} {45}},\ \bibinfo
  {pages} {57} (\bibinfo {year} {2000})}\BibitemShut {NoStop}%
\bibitem [{\citenamefont {Nishii}\ \emph {et~al.}(2002)\citenamefont {Nishii},
  \citenamefont {Kinugasa}, \citenamefont {Nii},\ and\ \citenamefont
  {Takahashi}}]{nishii_transport_2002}%
  \BibitemOpen
  \bibfield  {author} {\bibinfo {author} {\bibfnamefont {Y.}~\bibnamefont
  {Nishii}}, \bibinfo {author} {\bibfnamefont {T.}~\bibnamefont {Kinugasa}},
  \bibinfo {author} {\bibfnamefont {S.}~\bibnamefont {Nii}}, \ and\ \bibinfo
  {author} {\bibfnamefont {K.}~\bibnamefont {Takahashi}},\ }\href {\doibase
  10.1016/S0376-7388(01)00486-0} {\bibfield  {journal} {\bibinfo  {journal} {J.
  Memb. Sci.}\ }\textbf {\bibinfo {volume} {195}},\ \bibinfo {pages} {11}
  (\bibinfo {year} {2002})}\BibitemShut {NoStop}%
\bibitem [{\citenamefont {Nucci}, \citenamefont {Valentine},\ and\
  \citenamefont {Wand}(2014)}]{nucci_high-resolution_2014}%
  \BibitemOpen
  \bibfield  {author} {\bibinfo {author} {\bibfnamefont {N.~V.}\ \bibnamefont
  {Nucci}}, \bibinfo {author} {\bibfnamefont {K.~G.}\ \bibnamefont
  {Valentine}}, \ and\ \bibinfo {author} {\bibfnamefont {A.~J.}\ \bibnamefont
  {Wand}},\ }\href {\doibase 10.1016/j.jmr.2013.10.006} {\bibfield  {journal}
  {\bibinfo  {journal} {J. Magn. Reson.}\ }\bibinfo {series} {A special
  “{JMR} Perspectives” issue: Foresights in Biomolecular Solution-State
  {NMR} Spectroscopy – From Spin Gymnastics to Structure and Dynamics},\
  \textbf {\bibinfo {volume} {241}},\ \bibinfo {pages} {137} (\bibinfo {year}
  {2014})}\BibitemShut {NoStop}%
\bibitem [{\citenamefont {Dodevski}\ \emph {et~al.}(2014)\citenamefont
  {Dodevski}, \citenamefont {Nucci}, \citenamefont {Valentine}, \citenamefont
  {Sidhu}, \citenamefont {O’Brien}, \citenamefont {Pardi},\ and\
  \citenamefont {Wand}}]{dodevski_optimized_2014}%
  \BibitemOpen
  \bibfield  {author} {\bibinfo {author} {\bibfnamefont {I.}~\bibnamefont
  {Dodevski}}, \bibinfo {author} {\bibfnamefont {N.~V.}\ \bibnamefont {Nucci}},
  \bibinfo {author} {\bibfnamefont {K.~G.}\ \bibnamefont {Valentine}}, \bibinfo
  {author} {\bibfnamefont {G.~K.}\ \bibnamefont {Sidhu}}, \bibinfo {author}
  {\bibfnamefont {E.~S.}\ \bibnamefont {O’Brien}}, \bibinfo {author}
  {\bibfnamefont {A.}~\bibnamefont {Pardi}}, \ and\ \bibinfo {author}
  {\bibfnamefont {A.~J.}\ \bibnamefont {Wand}},\ }\href {\doibase
  10.1021/ja410716w} {\bibfield  {journal} {\bibinfo  {journal} {J. Am. Chem.
  Soc.}\ }\textbf {\bibinfo {volume} {136}},\ \bibinfo {pages} {3465} (\bibinfo
  {year} {2014})}\BibitemShut {NoStop}%
\bibitem [{\citenamefont {Galamba}(2013)}]{galamba_waters_2013}%
  \BibitemOpen
  \bibfield  {author} {\bibinfo {author} {\bibfnamefont {N.}~\bibnamefont
  {Galamba}},\ }\href {\doibase 10.1021/jp310649n} {\bibfield  {journal}
  {\bibinfo  {journal} {J. Phys. Chem. B}\ }\textbf {\bibinfo {volume} {117}},\
  \bibinfo {pages} {2153} (\bibinfo {year} {2013})}\BibitemShut {NoStop}%
\bibitem [{\citenamefont {Galamba}(2014)}]{galamba_water_2014}%
  \BibitemOpen
  \bibfield  {author} {\bibinfo {author} {\bibfnamefont {N.}~\bibnamefont
  {Galamba}},\ }\href {\doibase 10.1021/jp500067a} {\bibfield  {journal}
  {\bibinfo  {journal} {J. Phys. Chem. B}\ }\textbf {\bibinfo {volume} {118}},\
  \bibinfo {pages} {4169} (\bibinfo {year} {2014})}\BibitemShut {NoStop}%
\bibitem [{\citenamefont {Schick}(1987)}]{schick1987nonionic}%
  \BibitemOpen
  \bibfield  {author} {\bibinfo {author} {\bibfnamefont {M.~J.}\ \bibnamefont
  {Schick}},\ }\href@noop {} {\emph {\bibinfo {title} {Nonionic surfactants:
  physical chemistry}}}\ (\bibinfo  {publisher} {CRC Press},\ \bibinfo {year}
  {1987})\BibitemShut {NoStop}%
\bibitem [{\citenamefont {Aranow}\ and\ \citenamefont
  {Witten}(1960)}]{aranow_environmental_1960}%
  \BibitemOpen
  \bibfield  {author} {\bibinfo {author} {\bibfnamefont {R.~H.}\ \bibnamefont
  {Aranow}}\ and\ \bibinfo {author} {\bibfnamefont {L.}~\bibnamefont
  {Witten}},\ }\href {\doibase 10.1021/j100840a010} {\bibfield  {journal}
  {\bibinfo  {journal} {J. Phys. Chem.}\ }\textbf {\bibinfo {volume} {64}},\
  \bibinfo {pages} {1643} (\bibinfo {year} {1960})}\BibitemShut {NoStop}%
\bibitem [{\citenamefont {Aranow}\ and\ \citenamefont
  {Witten}(1965)}]{aranow_additional_1965}%
  \BibitemOpen
  \bibfield  {author} {\bibinfo {author} {\bibfnamefont {R.~H.}\ \bibnamefont
  {Aranow}}\ and\ \bibinfo {author} {\bibfnamefont {L.}~\bibnamefont
  {Witten}},\ }\href {\doibase 10.1063/1.1696950} {\bibfield  {journal}
  {\bibinfo  {journal} {J. Chem. Phys.}\ }\textbf {\bibinfo {volume} {43}},\
  \bibinfo {pages} {1436} (\bibinfo {year} {1965})}\BibitemShut {NoStop}%
\bibitem [{\citenamefont {Shiraga}\ \emph {et~al.}(2014)\citenamefont
  {Shiraga}, \citenamefont {Suzuki}, \citenamefont {Kondo},\ and\ \citenamefont
  {Ogawa}}]{shiraga_hydration_2014}%
  \BibitemOpen
  \bibfield  {author} {\bibinfo {author} {\bibfnamefont {K.}~\bibnamefont
  {Shiraga}}, \bibinfo {author} {\bibfnamefont {T.}~\bibnamefont {Suzuki}},
  \bibinfo {author} {\bibfnamefont {N.}~\bibnamefont {Kondo}}, \ and\ \bibinfo
  {author} {\bibfnamefont {Y.}~\bibnamefont {Ogawa}},\ }\href {\doibase
  10.1063/1.4903544} {\bibfield  {journal} {\bibinfo  {journal} {J. Chem.
  Phys.}\ }\textbf {\bibinfo {volume} {141}},\ \bibinfo {pages} {235103}
  (\bibinfo {year} {2014})}\BibitemShut {NoStop}%
\bibitem [{\citenamefont {Frank}\ and\ \citenamefont
  {Evans}(1945)}]{frank_free_1945}%
  \BibitemOpen
  \bibfield  {author} {\bibinfo {author} {\bibfnamefont {H.~S.}\ \bibnamefont
  {Frank}}\ and\ \bibinfo {author} {\bibfnamefont {M.~W.}\ \bibnamefont
  {Evans}},\ }\href {\doibase 10.1063/1.1723985} {\bibfield  {journal}
  {\bibinfo  {journal} {J. Chem. Phys.}\ }\textbf {\bibinfo {volume} {13}},\
  \bibinfo {pages} {507} (\bibinfo {year} {1945})}\BibitemShut {NoStop}%
\bibitem [{\citenamefont {Eicke}\ and\ \citenamefont
  {Christen}(1978)}]{Eicke1978}%
  \BibitemOpen
  \bibfield  {author} {\bibinfo {author} {\bibfnamefont {H.-F.}\ \bibnamefont
  {Eicke}}\ and\ \bibinfo {author} {\bibfnamefont {H.}~\bibnamefont
  {Christen}},\ }\href {\doibase 10.1002/hlca.19780610631} {\bibfield
  {journal} {\bibinfo  {journal} {Helv. Chim. Acta}\ }\textbf {\bibinfo
  {volume} {61}},\ \bibinfo {pages} {2258} (\bibinfo {year}
  {1978})}\BibitemShut {NoStop}%
\bibitem [{\citenamefont {Van~Dijk}\ \emph {et~al.}(1989)\citenamefont
  {Van~Dijk}, \citenamefont {Joosten}, \citenamefont {Levine},\ and\
  \citenamefont {Bedeaux}}]{VanDijk1989}%
  \BibitemOpen
  \bibfield  {author} {\bibinfo {author} {\bibfnamefont {M.~A.}\ \bibnamefont
  {Van~Dijk}}, \bibinfo {author} {\bibfnamefont {J.~G.~H.}\ \bibnamefont
  {Joosten}}, \bibinfo {author} {\bibfnamefont {Y.~K.}\ \bibnamefont {Levine}},
  \ and\ \bibinfo {author} {\bibfnamefont {D.}~\bibnamefont {Bedeaux}},\ }\href
  {\doibase 10.1021/j100343a054} {\bibfield  {journal} {\bibinfo  {journal}
  {The Journal of Physical Chemistry}\ }\textbf {\bibinfo {volume} {93}},\
  \bibinfo {pages} {2506} (\bibinfo {year} {1989})},\ \Eprint
  {http://arxiv.org/abs/https://doi.org/10.1021/j100343a054}
  {https://doi.org/10.1021/j100343a054} \BibitemShut {NoStop}%
\bibitem [{\citenamefont {Fathi}\ \emph {et~al.}(2012)\citenamefont {Fathi},
  \citenamefont {Kelly}, \citenamefont {Vasquez},\ and\ \citenamefont
  {Graeve}}]{Fathi2012}%
  \BibitemOpen
  \bibfield  {author} {\bibinfo {author} {\bibfnamefont {H.}~\bibnamefont
  {Fathi}}, \bibinfo {author} {\bibfnamefont {J.~P.}\ \bibnamefont {Kelly}},
  \bibinfo {author} {\bibfnamefont {V.~R.}\ \bibnamefont {Vasquez}}, \ and\
  \bibinfo {author} {\bibfnamefont {O.~A.}\ \bibnamefont {Graeve}},\ }\href
  {\doibase 10.1021/la300586f} {\bibfield  {journal} {\bibinfo  {journal}
  {Langmuir}\ }\textbf {\bibinfo {volume} {28}},\ \bibinfo {pages} {9267}
  (\bibinfo {year} {2012})},\ \bibinfo {note} {pMID: 22642604},\ \Eprint
  {http://arxiv.org/abs/https://doi.org/10.1021/la300586f}
  {https://doi.org/10.1021/la300586f} \BibitemShut {NoStop}%
\bibitem [{\citenamefont {Chowdhary}\ and\ \citenamefont
  {Ladanyi}(2009)}]{chowdhary_molecular_2009}%
  \BibitemOpen
  \bibfield  {author} {\bibinfo {author} {\bibfnamefont {J.}~\bibnamefont
  {Chowdhary}}\ and\ \bibinfo {author} {\bibfnamefont {B.~M.}\ \bibnamefont
  {Ladanyi}},\ }\href {\doibase 10.1021/jp906915q} {\bibfield  {journal}
  {\bibinfo  {journal} {J. Phys. Chem. B}\ }\textbf {\bibinfo {volume} {113}},\
  \bibinfo {pages} {15029} (\bibinfo {year} {2009})}\BibitemShut {NoStop}%
\bibitem [{\citenamefont {Eskici}\ and\ \citenamefont
  {Axelsen}(2016)}]{Eskici2016}%
  \BibitemOpen
  \bibfield  {author} {\bibinfo {author} {\bibfnamefont {G.}~\bibnamefont
  {Eskici}}\ and\ \bibinfo {author} {\bibfnamefont {P.~H.}\ \bibnamefont
  {Axelsen}},\ }\href {\doibase 10.1021/acs.jpcb.6b06420} {\bibfield  {journal}
  {\bibinfo  {journal} {The Journal of Physical Chemistry B}\ }\textbf
  {\bibinfo {volume} {120}},\ \bibinfo {pages} {11337} (\bibinfo {year}
  {2016})}\BibitemShut {NoStop}%
\bibitem [{\citenamefont {Faeder}\ and\ \citenamefont
  {Ladanyi}(2000)}]{faeder_molecular_2000}%
  \BibitemOpen
  \bibfield  {author} {\bibinfo {author} {\bibfnamefont {J.}~\bibnamefont
  {Faeder}}\ and\ \bibinfo {author} {\bibfnamefont {B.~M.}\ \bibnamefont
  {Ladanyi}},\ }\href {\doibase 10.1021/jp993076u} {\bibfield  {journal}
  {\bibinfo  {journal} {J. Phys. Chem. B}\ }\textbf {\bibinfo {volume} {104}},\
  \bibinfo {pages} {1033} (\bibinfo {year} {2000})}\BibitemShut {NoStop}%
\bibitem [{\citenamefont {Faeder}\ and\ \citenamefont
  {Ladanyi}(2001)}]{faeder_solvation_2001}%
  \BibitemOpen
  \bibfield  {author} {\bibinfo {author} {\bibfnamefont {J.}~\bibnamefont
  {Faeder}}\ and\ \bibinfo {author} {\bibfnamefont {B.~M.}\ \bibnamefont
  {Ladanyi}},\ }\href@noop {} {\bibfield  {journal} {\bibinfo  {journal} {J.
  Phys. Chem. B}\ }\textbf {\bibinfo {volume} {105}},\ \bibinfo {pages} {11148}
  (\bibinfo {year} {2001})}\BibitemShut {NoStop}%
\bibitem [{\citenamefont {Chowdhary}\ and\ \citenamefont
  {Ladanyi}(2011)}]{chowdhary_molecular_2011}%
  \BibitemOpen
  \bibfield  {author} {\bibinfo {author} {\bibfnamefont {J.}~\bibnamefont
  {Chowdhary}}\ and\ \bibinfo {author} {\bibfnamefont {B.~M.}\ \bibnamefont
  {Ladanyi}},\ }\href {\doibase 10.1021/jp201866t} {\bibfield  {journal}
  {\bibinfo  {journal} {J. Phys. Chem. A}\ }\textbf {\bibinfo {volume} {115}},\
  \bibinfo {pages} {6306} (\bibinfo {year} {2011})},\ \bibinfo {note}
  {00017}\BibitemShut {NoStop}%
\bibitem [{\citenamefont {Graeve}\ \emph {et~al.}(2013)\citenamefont {Graeve},
  \citenamefont {Fathi}, \citenamefont {Kelly}, \citenamefont {Saterlie},
  \citenamefont {Sinha}, \citenamefont {Rojas-George}, \citenamefont
  {Kanakala}, \citenamefont {Brown},\ and\ \citenamefont {Lopez}}]{Graeve2013}%
  \BibitemOpen
  \bibfield  {author} {\bibinfo {author} {\bibfnamefont {O.~A.}\ \bibnamefont
  {Graeve}}, \bibinfo {author} {\bibfnamefont {H.}~\bibnamefont {Fathi}},
  \bibinfo {author} {\bibfnamefont {J.~P.}\ \bibnamefont {Kelly}}, \bibinfo
  {author} {\bibfnamefont {M.~S.}\ \bibnamefont {Saterlie}}, \bibinfo {author}
  {\bibfnamefont {K.}~\bibnamefont {Sinha}}, \bibinfo {author} {\bibfnamefont
  {G.}~\bibnamefont {Rojas-George}}, \bibinfo {author} {\bibfnamefont
  {R.}~\bibnamefont {Kanakala}}, \bibinfo {author} {\bibfnamefont {D.~R.}\
  \bibnamefont {Brown}}, \ and\ \bibinfo {author} {\bibfnamefont {E.~A.}\
  \bibnamefont {Lopez}},\ }\href {\doibase
  https://doi.org/10.1016/j.jcis.2013.07.003} {\bibfield  {journal} {\bibinfo
  {journal} {Journal of Colloid and Interface Science}\ }\textbf {\bibinfo
  {volume} {407}},\ \bibinfo {pages} {302 } (\bibinfo {year}
  {2013})}\BibitemShut {NoStop}%
\bibitem [{\citenamefont {Mukherjee}, \citenamefont {Chowdhury},\ and\
  \citenamefont {Gai}(2006)}]{Mukherjee2006}%
  \BibitemOpen
  \bibfield  {author} {\bibinfo {author} {\bibfnamefont {S.}~\bibnamefont
  {Mukherjee}}, \bibinfo {author} {\bibfnamefont {P.}~\bibnamefont
  {Chowdhury}}, \ and\ \bibinfo {author} {\bibfnamefont {F.}~\bibnamefont
  {Gai}},\ }\href {\doibase 10.1021/jp062362k} {\bibfield  {journal} {\bibinfo
  {journal} {The Journal of Physical Chemistry B}\ }\textbf {\bibinfo {volume}
  {110}},\ \bibinfo {pages} {11615} (\bibinfo {year} {2006})},\ \bibinfo {note}
  {pMID: 16800453},\ \Eprint
  {http://arxiv.org/abs/https://doi.org/10.1021/jp062362k}
  {https://doi.org/10.1021/jp062362k} \BibitemShut {NoStop}%
\bibitem [{\citenamefont {Mukherjee}\ \emph {et~al.}(2007)\citenamefont
  {Mukherjee}, \citenamefont {Chowdhury}, \citenamefont {DeGrado},\ and\
  \citenamefont {Gai}}]{Mukherjee2007}%
  \BibitemOpen
  \bibfield  {author} {\bibinfo {author} {\bibfnamefont {S.}~\bibnamefont
  {Mukherjee}}, \bibinfo {author} {\bibfnamefont {P.}~\bibnamefont
  {Chowdhury}}, \bibinfo {author} {\bibfnamefont {W.~F.}\ \bibnamefont
  {DeGrado}}, \ and\ \bibinfo {author} {\bibfnamefont {F.}~\bibnamefont
  {Gai}},\ }\href {\doibase 10.1021/la701686g} {\bibfield  {journal} {\bibinfo
  {journal} {Langmuir}\ }\textbf {\bibinfo {volume} {23}},\ \bibinfo {pages}
  {11174} (\bibinfo {year} {2007})},\ \bibinfo {note} {pMID: 17910485},\
  \Eprint {http://arxiv.org/abs/https://doi.org/10.1021/la701686g}
  {https://doi.org/10.1021/la701686g} \BibitemShut {NoStop}%
\bibitem [{\citenamefont {Mukherjee}, \citenamefont {Moulik},\ and\
  \citenamefont {Mukherjee}(1993)}]{mukherjee1993thermodynamics}%
  \BibitemOpen
  \bibfield  {author} {\bibinfo {author} {\bibfnamefont {K.}~\bibnamefont
  {Mukherjee}}, \bibinfo {author} {\bibfnamefont {S.}~\bibnamefont {Moulik}}, \
  and\ \bibinfo {author} {\bibfnamefont {D.}~\bibnamefont {Mukherjee}},\
  }\href@noop {} {\bibfield  {journal} {\bibinfo  {journal} {Langmuir}\
  }\textbf {\bibinfo {volume} {9}},\ \bibinfo {pages} {1727} (\bibinfo {year}
  {1993})}\BibitemShut {NoStop}%
\bibitem [{\citenamefont {Ruckenstein}\ and\ \citenamefont
  {Nagarajan}(1980)}]{ruckenstein1980aggregation}%
  \BibitemOpen
  \bibfield  {author} {\bibinfo {author} {\bibfnamefont {E.}~\bibnamefont
  {Ruckenstein}}\ and\ \bibinfo {author} {\bibfnamefont {R.}~\bibnamefont
  {Nagarajan}},\ }\href@noop {} {\bibfield  {journal} {\bibinfo  {journal} {J.
  Phys. Chem.}\ }\textbf {\bibinfo {volume} {84}},\ \bibinfo {pages} {1349}
  (\bibinfo {year} {1980})}\BibitemShut {NoStop}%
\bibitem [{\citenamefont {Ekwall}, \citenamefont {Mandell},\ and\ \citenamefont
  {Fontell}(1970)}]{ekwall1970some}%
  \BibitemOpen
  \bibfield  {author} {\bibinfo {author} {\bibfnamefont {P.}~\bibnamefont
  {Ekwall}}, \bibinfo {author} {\bibfnamefont {L.}~\bibnamefont {Mandell}}, \
  and\ \bibinfo {author} {\bibfnamefont {K.}~\bibnamefont {Fontell}},\
  }\href@noop {} {\bibfield  {journal} {\bibinfo  {journal} {J. Colloid
  Interface Sci.}\ }\textbf {\bibinfo {volume} {33}},\ \bibinfo {pages} {215}
  (\bibinfo {year} {1970})}\BibitemShut {NoStop}%
\bibitem [{\citenamefont {Wootton}, \citenamefont {Picavez},\ and\
  \citenamefont {Harrowell}(2008)}]{Wootton2008}%
  \BibitemOpen
  \bibfield  {author} {\bibinfo {author} {\bibfnamefont {A.}~\bibnamefont
  {Wootton}}, \bibinfo {author} {\bibfnamefont {F.}~\bibnamefont {Picavez}}, \
  and\ \bibinfo {author} {\bibfnamefont {P.}~\bibnamefont {Harrowell}},\ }\href
  {\doibase 10.1063/1.2897802} {\bibfield  {journal} {\bibinfo  {journal} {AIP
  Conference Proceedings}\ }\textbf {\bibinfo {volume} {982}},\ \bibinfo
  {pages} {289} (\bibinfo {year} {2008})},\ \Eprint
  {http://arxiv.org/abs/https://aip.scitation.org/doi/pdf/10.1063/1.2897802}
  {https://aip.scitation.org/doi/pdf/10.1063/1.2897802} \BibitemShut {NoStop}%
\bibitem [{\citenamefont {Tanaka}(1992)}]{TANAKA1992}%
  \BibitemOpen
  \bibfield  {author} {\bibinfo {author} {\bibfnamefont {R.}~\bibnamefont
  {Tanaka}},\ }\href {\doibase 10.5650/jos1956.41.82} {\bibfield  {journal}
  {\bibinfo  {journal} {Journal of Japan Oil Chemists' Society}\ }\textbf
  {\bibinfo {volume} {41}},\ \bibinfo {pages} {82} (\bibinfo {year}
  {1992})}\BibitemShut {NoStop}%
\bibitem [{\citenamefont {Tanaka}\ \emph {et~al.}(2005)\citenamefont {Tanaka},
  \citenamefont {Yokoyama}, \citenamefont {Sameshima},\ and\ \citenamefont
  {Kawase}}]{Tanaka2005}%
  \BibitemOpen
  \bibfield  {author} {\bibinfo {author} {\bibfnamefont {R.}~\bibnamefont
  {Tanaka}}, \bibinfo {author} {\bibfnamefont {T.}~\bibnamefont {Yokoyama}},
  \bibinfo {author} {\bibfnamefont {K.}~\bibnamefont {Sameshima}}, \ and\
  \bibinfo {author} {\bibfnamefont {T.}~\bibnamefont {Kawase}},\ }\href
  {\doibase 10.1246/bcsj.78.599} {\bibfield  {journal} {\bibinfo  {journal}
  {Bull. Chem. Soc. Jpn.}\ }\textbf {\bibinfo {volume} {78}},\ \bibinfo {pages}
  {599} (\bibinfo {year} {2005})}\BibitemShut {NoStop}%
\bibitem [{\citenamefont {Sameshima}\ \emph {et~al.}(2006)\citenamefont
  {Sameshima}, \citenamefont {Tanaka}, \citenamefont {Igarashi},\ and\
  \citenamefont {Ooshima}}]{Sameshima2006}%
  \BibitemOpen
  \bibfield  {author} {\bibinfo {author} {\bibfnamefont {K.}~\bibnamefont
  {Sameshima}}, \bibinfo {author} {\bibfnamefont {R.}~\bibnamefont {Tanaka}},
  \bibinfo {author} {\bibfnamefont {K.}~\bibnamefont {Igarashi}}, \ and\
  \bibinfo {author} {\bibfnamefont {H.}~\bibnamefont {Ooshima}},\ }\href
  {\doibase 10.1016/j.jct.2005.07.021} {\bibfield  {journal} {\bibinfo
  {journal} {J. Chem. Thermodyn.}\ }\textbf {\bibinfo {volume} {38}},\ \bibinfo
  {pages} {662} (\bibinfo {year} {2006})}\BibitemShut {NoStop}%
\bibitem [{\citenamefont {Chen}(1986)}]{Chen1986}%
  \BibitemOpen
  \bibfield  {author} {\bibinfo {author} {\bibfnamefont {S.~H.}\ \bibnamefont
  {Chen}},\ }\href {\doibase 10.1146/annurev.pc.37.100186.002031} {\bibfield
  {journal} {\bibinfo  {journal} {Ann. Rev. Phys. Chem.}\ }\textbf {\bibinfo
  {volume} {37}},\ \bibinfo {pages} {351} (\bibinfo {year} {1986})}\BibitemShut
  {NoStop}%
\bibitem [{\citenamefont {Muto}\ and\ \citenamefont {Meguro}(1973)}]{Muto1973}%
  \BibitemOpen
  \bibfield  {author} {\bibinfo {author} {\bibfnamefont {S.}~\bibnamefont
  {Muto}}\ and\ \bibinfo {author} {\bibfnamefont {K.}~\bibnamefont {Meguro}},\
  }\href {\doibase 10.1246/bcsj.46.1316} {\bibfield  {journal} {\bibinfo
  {journal} {Bull. Chem. Soc. Jpn.}\ }\textbf {\bibinfo {volume} {46}},\
  \bibinfo {pages} {1316} (\bibinfo {year} {1973})}\BibitemShut {NoStop}%
\bibitem [{\citenamefont {Kon-no}\ and\ \citenamefont
  {Kitahara}(1971)}]{Kon-no1971}%
  \BibitemOpen
  \bibfield  {author} {\bibinfo {author} {\bibfnamefont {K.}~\bibnamefont
  {Kon-no}}\ and\ \bibinfo {author} {\bibfnamefont {A.}~\bibnamefont
  {Kitahara}},\ }\href {\doibase 10.1016/0021-9797(71)90222-0} {\bibfield
  {journal} {\bibinfo  {journal} {J. Colloid Interface Sci.}\ }\textbf
  {\bibinfo {volume} {35}},\ \bibinfo {pages} {636} (\bibinfo {year}
  {1971})}\BibitemShut {NoStop}%
\bibitem [{\citenamefont {Kitahara}, \citenamefont {Kobayashi},\ and\
  \citenamefont {Tachibana}(1962)}]{Kitahara1962}%
  \BibitemOpen
  \bibfield  {author} {\bibinfo {author} {\bibfnamefont {A.}~\bibnamefont
  {Kitahara}}, \bibinfo {author} {\bibfnamefont {T.}~\bibnamefont {Kobayashi}},
  \ and\ \bibinfo {author} {\bibfnamefont {T.}~\bibnamefont {Tachibana}},\
  }\href {\doibase 10.1021/j100808a510} {\bibfield  {journal} {\bibinfo
  {journal} {The J. Phys. Chem.}\ }\textbf {\bibinfo {volume} {66}},\ \bibinfo
  {pages} {363} (\bibinfo {year} {1962})}\BibitemShut {NoStop}%
\bibitem [{\citenamefont {Frank}\ and\ \citenamefont
  {Zografi}(1969)}]{Frank1969}%
  \BibitemOpen
  \bibfield  {author} {\bibinfo {author} {\bibfnamefont {S.~G.}\ \bibnamefont
  {Frank}}\ and\ \bibinfo {author} {\bibfnamefont {G.}~\bibnamefont
  {Zografi}},\ }\href {\doibase 10.1002/jps.2600580820} {\bibfield  {journal}
  {\bibinfo  {journal} {J. Pharm. Sci.}\ }\textbf {\bibinfo {volume} {58}},\
  \bibinfo {pages} {993} (\bibinfo {year} {1969})}\BibitemShut {NoStop}%
\bibitem [{\citenamefont {Smith}\ \emph {et~al.}(2013)\citenamefont {Smith},
  \citenamefont {Brown}, \citenamefont {Rogers},\ and\ \citenamefont
  {Eastoe}}]{Smith2013}%
  \BibitemOpen
  \bibfield  {author} {\bibinfo {author} {\bibfnamefont {G.~N.}\ \bibnamefont
  {Smith}}, \bibinfo {author} {\bibfnamefont {P.}~\bibnamefont {Brown}},
  \bibinfo {author} {\bibfnamefont {S.~E.}\ \bibnamefont {Rogers}}, \ and\
  \bibinfo {author} {\bibfnamefont {J.}~\bibnamefont {Eastoe}},\ }\href
  {\doibase 10.1021/la400117s} {\bibfield  {journal} {\bibinfo  {journal}
  {Langmuir}\ }\textbf {\bibinfo {volume} {29}},\ \bibinfo {pages} {3252}
  (\bibinfo {year} {2013})}\BibitemShut {NoStop}%
\bibitem [{\citenamefont {Smith}\ \emph {et~al.}(2016)\citenamefont {Smith},
  \citenamefont {Brown}, \citenamefont {James}, \citenamefont {Rogers},\ and\
  \citenamefont {Eastoe}}]{smith2016effect}%
  \BibitemOpen
  \bibfield  {author} {\bibinfo {author} {\bibfnamefont {G.~N.}\ \bibnamefont
  {Smith}}, \bibinfo {author} {\bibfnamefont {P.}~\bibnamefont {Brown}},
  \bibinfo {author} {\bibfnamefont {C.}~\bibnamefont {James}}, \bibinfo
  {author} {\bibfnamefont {S.~E.}\ \bibnamefont {Rogers}}, \ and\ \bibinfo
  {author} {\bibfnamefont {J.}~\bibnamefont {Eastoe}},\ }\href@noop {}
  {\bibfield  {journal} {\bibinfo  {journal} {Colloids Surf. A Physicochem Eng.
  Asp.}\ }\textbf {\bibinfo {volume} {494}},\ \bibinfo {pages} {194} (\bibinfo
  {year} {2016})}\BibitemShut {NoStop}%
\bibitem [{\citenamefont {Hollamby}\ \emph {et~al.}(2008)\citenamefont
  {Hollamby}, \citenamefont {Tabor}, \citenamefont {Mutch}, \citenamefont
  {Trickett}, \citenamefont {Eastoe}, \citenamefont {Heenan},\ and\
  \citenamefont {Grillo}}]{hollamby_effect_2008}%
  \BibitemOpen
  \bibfield  {author} {\bibinfo {author} {\bibfnamefont {M.~J.}\ \bibnamefont
  {Hollamby}}, \bibinfo {author} {\bibfnamefont {R.}~\bibnamefont {Tabor}},
  \bibinfo {author} {\bibfnamefont {K.~J.}\ \bibnamefont {Mutch}}, \bibinfo
  {author} {\bibfnamefont {K.}~\bibnamefont {Trickett}}, \bibinfo {author}
  {\bibfnamefont {J.}~\bibnamefont {Eastoe}}, \bibinfo {author} {\bibfnamefont
  {R.~K.}\ \bibnamefont {Heenan}}, \ and\ \bibinfo {author} {\bibfnamefont
  {I.}~\bibnamefont {Grillo}},\ }\href {\doibase 10.1021/la8020854} {\bibfield
  {journal} {\bibinfo  {journal} {Langmuir}\ }\textbf {\bibinfo {volume}
  {24}},\ \bibinfo {pages} {12235} (\bibinfo {year} {2008})}\BibitemShut
  {NoStop}%
\bibitem [{\citenamefont {Christopher}\ and\ \citenamefont
  {Oxtoby}(2003)}]{christopher2003free}%
  \BibitemOpen
  \bibfield  {author} {\bibinfo {author} {\bibfnamefont {P.}~\bibnamefont
  {Christopher}}\ and\ \bibinfo {author} {\bibfnamefont {D.~W.}\ \bibnamefont
  {Oxtoby}},\ }\href@noop {} {\bibfield  {journal} {\bibinfo  {journal} {J.
  Phys. Chem.}\ }\textbf {\bibinfo {volume} {118}},\ \bibinfo {pages} {5665}
  (\bibinfo {year} {2003})}\BibitemShut {NoStop}%
\bibitem [{\citenamefont {Mohan}\ and\ \citenamefont
  {Kopelevich}(2008)}]{mohan2008multiscale}%
  \BibitemOpen
  \bibfield  {author} {\bibinfo {author} {\bibfnamefont {G.}~\bibnamefont
  {Mohan}}\ and\ \bibinfo {author} {\bibfnamefont {D.~I.}\ \bibnamefont
  {Kopelevich}},\ }\href@noop {} {\bibfield  {journal} {\bibinfo  {journal} {J.
  Phys. Chem.}\ }\textbf {\bibinfo {volume} {128}},\ \bibinfo {pages} {044905}
  (\bibinfo {year} {2008})}\BibitemShut {NoStop}%
\bibitem [{\citenamefont {Kindt}(2013)}]{Kindt2013}%
  \BibitemOpen
  \bibfield  {author} {\bibinfo {author} {\bibfnamefont {J.~T.}\ \bibnamefont
  {Kindt}},\ }\href {\doibase 10.1021/ct300686u} {\bibfield  {journal}
  {\bibinfo  {journal} {J. Chem. Theory Comput.}\ }\textbf {\bibinfo {volume}
  {9}},\ \bibinfo {pages} {147} (\bibinfo {year} {2013})}\BibitemShut {NoStop}%
\bibitem [{\citenamefont {Ben-Shaul}\ and\ \citenamefont
  {Gelbart}(1982)}]{Ben-Shaul1982}%
  \BibitemOpen
  \bibfield  {author} {\bibinfo {author} {\bibfnamefont {A.}~\bibnamefont
  {Ben-Shaul}}\ and\ \bibinfo {author} {\bibfnamefont {W.~M.}\ \bibnamefont
  {Gelbart}},\ }\href {\doibase 10.1021/j100392a004} {\bibfield  {journal}
  {\bibinfo  {journal} {The J. Phys. Chem.}\ }\textbf {\bibinfo {volume}
  {86}},\ \bibinfo {pages} {316} (\bibinfo {year} {1982})}\BibitemShut
  {NoStop}%
\bibitem [{\citenamefont {Kinoshita}\ and\ \citenamefont
  {Sugai}(1999)}]{kinoshita1999new}%
  \BibitemOpen
  \bibfield  {author} {\bibinfo {author} {\bibfnamefont {M.}~\bibnamefont
  {Kinoshita}}\ and\ \bibinfo {author} {\bibfnamefont {Y.}~\bibnamefont
  {Sugai}},\ }\href@noop {} {\bibfield  {journal} {\bibinfo  {journal} {Chem.
  Phys. letters}\ }\textbf {\bibinfo {volume} {313}},\ \bibinfo {pages} {685}
  (\bibinfo {year} {1999})}\BibitemShut {NoStop}%
\bibitem [{\citenamefont {Kinoshita}\ and\ \citenamefont
  {Sugai}(2002)}]{kinoshita2002methodology}%
  \BibitemOpen
  \bibfield  {author} {\bibinfo {author} {\bibfnamefont {M.}~\bibnamefont
  {Kinoshita}}\ and\ \bibinfo {author} {\bibfnamefont {Y.}~\bibnamefont
  {Sugai}},\ }\href@noop {} {\bibfield  {journal} {\bibinfo  {journal} {J.
  Comput. Chem.}\ }\textbf {\bibinfo {volume} {23}},\ \bibinfo {pages} {1445}
  (\bibinfo {year} {2002})}\BibitemShut {NoStop}%
\bibitem [{\citenamefont {Yoshii}\ and\ \citenamefont
  {Okazaki}(2006{\natexlab{a}})}]{Yoshii2006}%
  \BibitemOpen
  \bibfield  {author} {\bibinfo {author} {\bibfnamefont {N.}~\bibnamefont
  {Yoshii}}\ and\ \bibinfo {author} {\bibfnamefont {S.}~\bibnamefont
  {Okazaki}},\ }\href {\doibase 10.1016/j.cplett.2006.05.004} {\bibfield
  {journal} {\bibinfo  {journal} {Chem. Phys, Lett.}\ }\textbf {\bibinfo
  {volume} {425}},\ \bibinfo {pages} {58} (\bibinfo {year}
  {2006}{\natexlab{a}})}\BibitemShut {NoStop}%
\bibitem [{\citenamefont {Yoshii}\ and\ \citenamefont
  {Okazaki}(2006{\natexlab{b}})}]{Yoshii2006a}%
  \BibitemOpen
  \bibfield  {author} {\bibinfo {author} {\bibfnamefont {N.}~\bibnamefont
  {Yoshii}}\ and\ \bibinfo {author} {\bibfnamefont {S.}~\bibnamefont
  {Okazaki}},\ }\href {\doibase 10.1016/j.cplett.2006.05.038} {\bibfield
  {journal} {\bibinfo  {journal} {Chem. Phys, Lett.}\ }\textbf {\bibinfo
  {volume} {426}},\ \bibinfo {pages} {66} (\bibinfo {year}
  {2006}{\natexlab{b}})}\BibitemShut {NoStop}%
\bibitem [{\citenamefont {Yoshii}, \citenamefont {Iwahashi},\ and\
  \citenamefont {Okazaki}(2006)}]{Yoshii2006b}%
  \BibitemOpen
  \bibfield  {author} {\bibinfo {author} {\bibfnamefont {N.}~\bibnamefont
  {Yoshii}}, \bibinfo {author} {\bibfnamefont {K.}~\bibnamefont {Iwahashi}}, \
  and\ \bibinfo {author} {\bibfnamefont {S.}~\bibnamefont {Okazaki}},\ }\href
  {\doibase 10.1063/1.2179074} {\bibfield  {journal} {\bibinfo  {journal} {J.
  Chem. Phys.}\ }\textbf {\bibinfo {volume} {124}},\ \bibinfo {pages} {184901}
  (\bibinfo {year} {2006})}\BibitemShut {NoStop}%
\bibitem [{\citenamefont {Yoshii}\ and\ \citenamefont
  {Okazaki}(2007)}]{Yoshii}%
  \BibitemOpen
  \bibfield  {author} {\bibinfo {author} {\bibfnamefont {N.}~\bibnamefont
  {Yoshii}}\ and\ \bibinfo {author} {\bibfnamefont {S.}~\bibnamefont
  {Okazaki}},\ }\href
  {http://www.icmp.lviv.ua/journal/zbirnyk.52/009/art09.pdf} {\bibfield
  {journal} {\bibinfo  {journal} {Cond. Mat. Phys}\ }\textbf {\bibinfo {volume}
  {10}},\ \bibinfo {pages} {573} (\bibinfo {year} {2007})}\BibitemShut
  {NoStop}%
\bibitem [{\citenamefont {Matubayasi}\ and\ \citenamefont
  {Nakahara}(2000)}]{Matubayasi2000}%
  \BibitemOpen
  \bibfield  {author} {\bibinfo {author} {\bibfnamefont {N.}~\bibnamefont
  {Matubayasi}}\ and\ \bibinfo {author} {\bibfnamefont {M.}~\bibnamefont
  {Nakahara}},\ }\href {\doibase 10.1063/1.1309013} {\bibfield  {journal}
  {\bibinfo  {journal} {J. Chem. Phys.}\ }\textbf {\bibinfo {volume} {113}},\
  \bibinfo {pages} {6070} (\bibinfo {year} {2000})}\BibitemShut {NoStop}%
\bibitem [{\citenamefont {Matubayasi}\ and\ \citenamefont
  {Nakahara}(2002)}]{Matubayasi2002}%
  \BibitemOpen
  \bibfield  {author} {\bibinfo {author} {\bibfnamefont {N.}~\bibnamefont
  {Matubayasi}}\ and\ \bibinfo {author} {\bibfnamefont {M.}~\bibnamefont
  {Nakahara}},\ }\href {\doibase 10.1063/1.1495850} {\bibfield  {journal}
  {\bibinfo  {journal} {J. Chem. Phys.}\ }\textbf {\bibinfo {volume} {117}},\
  \bibinfo {pages} {3605} (\bibinfo {year} {2002})}\BibitemShut {NoStop}%
\bibitem [{\citenamefont {Matubayasi}\ and\ \citenamefont
  {Nakahara}(2003)}]{Matubayasi2003}%
  \BibitemOpen
  \bibfield  {author} {\bibinfo {author} {\bibfnamefont {N.}~\bibnamefont
  {Matubayasi}}\ and\ \bibinfo {author} {\bibfnamefont {M.}~\bibnamefont
  {Nakahara}},\ }\href {\doibase 10.1063/1.1613938} {\bibfield  {journal}
  {\bibinfo  {journal} {J. Chem. Phys.}\ }\textbf {\bibinfo {volume} {119}},\
  \bibinfo {pages} {9686} (\bibinfo {year} {2003})}\BibitemShut {NoStop}%
\bibitem [{\citenamefont {Tanford}(1980)}]{tanford1980hydrophobic}%
  \BibitemOpen
  \bibfield  {author} {\bibinfo {author} {\bibfnamefont {C.}~\bibnamefont
  {Tanford}},\ }\href@noop {} {\emph {\bibinfo {title} {The Hydrophobic Effect:
  Formation of Micelles and Biological Membranes 2d Ed}}}\ (\bibinfo
  {publisher} {J. Wiley.},\ \bibinfo {year} {1980})\BibitemShut {NoStop}%
\bibitem [{\citenamefont {Puvvada}\ and\ \citenamefont
  {Blankschtein}(1990)}]{puvvada1990molecular}%
  \BibitemOpen
  \bibfield  {author} {\bibinfo {author} {\bibfnamefont {S.}~\bibnamefont
  {Puvvada}}\ and\ \bibinfo {author} {\bibfnamefont {D.}~\bibnamefont
  {Blankschtein}},\ }\href@noop {} {\bibfield  {journal} {\bibinfo  {journal}
  {J. Phys. Chem.}\ }\textbf {\bibinfo {volume} {92}},\ \bibinfo {pages} {3710}
  (\bibinfo {year} {1990})}\BibitemShut {NoStop}%
\bibitem [{\citenamefont {Zoeller}, \citenamefont {Lue},\ and\ \citenamefont
  {Blankschtein}(1997)}]{zoeller1997statistical}%
  \BibitemOpen
  \bibfield  {author} {\bibinfo {author} {\bibfnamefont {N.}~\bibnamefont
  {Zoeller}}, \bibinfo {author} {\bibfnamefont {L.}~\bibnamefont {Lue}}, \ and\
  \bibinfo {author} {\bibfnamefont {D.}~\bibnamefont {Blankschtein}},\
  }\href@noop {} {\bibfield  {journal} {\bibinfo  {journal} {Langmuir}\
  }\textbf {\bibinfo {volume} {13}},\ \bibinfo {pages} {5258} (\bibinfo {year}
  {1997})}\BibitemShut {NoStop}%
\bibitem [{\citenamefont {Kindt}()}]{kindtpersonal}%
  \BibitemOpen
  \bibfield  {author} {\bibinfo {author} {\bibfnamefont {J.~T.}\ \bibnamefont
  {Kindt}},\ }\href@noop {} {}\bibinfo {howpublished} {personal
  communication}\BibitemShut {NoStop}%
\bibitem [{\citenamefont {Frolov}(2015)}]{Frolov2015}%
  \BibitemOpen
  \bibfield  {author} {\bibinfo {author} {\bibfnamefont {A.~I.}\ \bibnamefont
  {Frolov}},\ }\href {\doibase 10.1021/acs.jctc.5b00172} {\bibfield  {journal}
  {\bibinfo  {journal} {J. Chem. Theory Comput.}\ }\textbf {\bibinfo {volume}
  {11}},\ \bibinfo {pages} {2245} (\bibinfo {year} {2015})}\BibitemShut
  {NoStop}%
\bibitem [{\citenamefont {Sakuraba}\ and\ \citenamefont
  {Matubayasi}(2014)}]{Sakuraba2014}%
  \BibitemOpen
  \bibfield  {author} {\bibinfo {author} {\bibfnamefont {S.}~\bibnamefont
  {Sakuraba}}\ and\ \bibinfo {author} {\bibfnamefont {N.}~\bibnamefont
  {Matubayasi}},\ }\href {\doibase 10.1002/jcc.23651} {\bibfield  {journal}
  {\bibinfo  {journal} {J. Comput. Chem.}\ }\textbf {\bibinfo {volume} {35}},\
  \bibinfo {pages} {1592} (\bibinfo {year} {2014})}\BibitemShut {NoStop}%
\bibitem [{\citenamefont {Mart{\'\i}nez}\ \emph {et~al.}(2009)\citenamefont
  {Mart{\'\i}nez}, \citenamefont {Andrade}, \citenamefont {Birgin},\ and\
  \citenamefont {Mart{\'\i}nez}}]{martinez2009packmol}%
  \BibitemOpen
  \bibfield  {author} {\bibinfo {author} {\bibfnamefont {L.}~\bibnamefont
  {Mart{\'\i}nez}}, \bibinfo {author} {\bibfnamefont {R.}~\bibnamefont
  {Andrade}}, \bibinfo {author} {\bibfnamefont {E.~G.}\ \bibnamefont {Birgin}},
  \ and\ \bibinfo {author} {\bibfnamefont {J.~M.}\ \bibnamefont
  {Mart{\'\i}nez}},\ }\href@noop {} {\bibfield  {journal} {\bibinfo  {journal}
  {J. Comput. Chem.}\ }\textbf {\bibinfo {volume} {30}},\ \bibinfo {pages}
  {2157} (\bibinfo {year} {2009})}\BibitemShut {NoStop}%
\bibitem [{\citenamefont {Abel}\ \emph {et~al.}(2004)\citenamefont {Abel},
  \citenamefont {Sterpone}, \citenamefont {Bandyopadhyay},\ and\ \citenamefont
  {Marchi}}]{Abel2004}%
  \BibitemOpen
  \bibfield  {author} {\bibinfo {author} {\bibfnamefont {S.}~\bibnamefont
  {Abel}}, \bibinfo {author} {\bibfnamefont {F.}~\bibnamefont {Sterpone}},
  \bibinfo {author} {\bibfnamefont {S.}~\bibnamefont {Bandyopadhyay}}, \ and\
  \bibinfo {author} {\bibfnamefont {M.}~\bibnamefont {Marchi}},\ }\href
  {\doibase 10.1021/jp047138e} {\bibfield  {journal} {\bibinfo  {journal} {J.
  Phys. Chem. B}\ }\textbf {\bibinfo {volume} {108}},\ \bibinfo {pages} {19458}
  (\bibinfo {year} {2004})}\BibitemShut {NoStop}%
\bibitem [{\citenamefont {Humphrey}, \citenamefont {Dalke},\ and\ \citenamefont
  {Schulten}(1996)}]{HUMP96}%
  \BibitemOpen
  \bibfield  {author} {\bibinfo {author} {\bibfnamefont {W.}~\bibnamefont
  {Humphrey}}, \bibinfo {author} {\bibfnamefont {A.}~\bibnamefont {Dalke}}, \
  and\ \bibinfo {author} {\bibfnamefont {K.}~\bibnamefont {Schulten}},\ }\href
  {\doibase 10.1016/0263-7855(96)00018-5} {\bibfield  {journal} {\bibinfo
  {journal} {J. Mol. Graph.}\ }\textbf {\bibinfo {volume} {14}},\ \bibinfo
  {pages} {33} (\bibinfo {year} {1996})}\BibitemShut {NoStop}%
\bibitem [{\citenamefont {Berendsen}, \citenamefont {van~der Spoel},\ and\
  \citenamefont {van Drunen}(1995)}]{Berendsen1995}%
  \BibitemOpen
  \bibfield  {author} {\bibinfo {author} {\bibfnamefont {H.}~\bibnamefont
  {Berendsen}}, \bibinfo {author} {\bibfnamefont {D.}~\bibnamefont {van~der
  Spoel}}, \ and\ \bibinfo {author} {\bibfnamefont {R.}~\bibnamefont {van
  Drunen}},\ }\href {\doibase 10.1016/0010-4655(95)00042-E} {\bibfield
  {journal} {\bibinfo  {journal} {Comput. Phys. Commun.}\ }\textbf {\bibinfo
  {volume} {91}},\ \bibinfo {pages} {43} (\bibinfo {year} {1995})}\BibitemShut
  {NoStop}%
\bibitem [{\citenamefont {P{\'{a}}ll}\ \emph {et~al.}(2015)\citenamefont
  {P{\'{a}}ll}, \citenamefont {Abraham}, \citenamefont {Kutzner}, \citenamefont
  {Hess},\ and\ \citenamefont {Lindahl}}]{Pall2015}%
  \BibitemOpen
  \bibfield  {author} {\bibinfo {author} {\bibfnamefont {S.}~\bibnamefont
  {P{\'{a}}ll}}, \bibinfo {author} {\bibfnamefont {M.~J.}\ \bibnamefont
  {Abraham}}, \bibinfo {author} {\bibfnamefont {C.}~\bibnamefont {Kutzner}},
  \bibinfo {author} {\bibfnamefont {B.}~\bibnamefont {Hess}}, \ and\ \bibinfo
  {author} {\bibfnamefont {E.}~\bibnamefont {Lindahl}},\ }in\ \href {\doibase
  10.1007/978-3-319-15976-8_1} {\emph {\bibinfo {booktitle} {International
  Conference on Exascale Applications and Software}}},\ \bibinfo {editor}
  {edited by\ \bibinfo {editor} {\bibfnamefont {S.}~\bibnamefont {Markidis}}\
  and\ \bibinfo {editor} {\bibfnamefont {E.}~\bibnamefont {Laure}}}\ (\bibinfo
  {publisher} {Springer International Publishing},\ \bibinfo {year} {2015})\
  pp.\ \bibinfo {pages} {3--27}\BibitemShut {NoStop}%
\bibitem [{\citenamefont {Abraham}\ \emph {et~al.}(2015)\citenamefont
  {Abraham}, \citenamefont {Murtola}, \citenamefont {Schulz}, \citenamefont
  {P{\'{a}}ll}, \citenamefont {Smith}, \citenamefont {Hess},\ and\
  \citenamefont {Lindahl}}]{abraham2015gromacs}%
  \BibitemOpen
  \bibfield  {author} {\bibinfo {author} {\bibfnamefont {M.~J.}\ \bibnamefont
  {Abraham}}, \bibinfo {author} {\bibfnamefont {T.}~\bibnamefont {Murtola}},
  \bibinfo {author} {\bibfnamefont {R.}~\bibnamefont {Schulz}}, \bibinfo
  {author} {\bibfnamefont {S.}~\bibnamefont {P{\'{a}}ll}}, \bibinfo {author}
  {\bibfnamefont {J.~C.}\ \bibnamefont {Smith}}, \bibinfo {author}
  {\bibfnamefont {B.}~\bibnamefont {Hess}}, \ and\ \bibinfo {author}
  {\bibfnamefont {E.}~\bibnamefont {Lindahl}},\ }\href@noop {} {\bibfield
  {journal} {\bibinfo  {journal} {SoftwareX}\ }\textbf {\bibinfo {volume}
  {1}},\ \bibinfo {pages} {19} (\bibinfo {year} {2015})}\BibitemShut {NoStop}%
\bibitem [{\citenamefont {{R Core Team}}(2015)}]{Rpackage2013}%
  \BibitemOpen
  \bibfield  {author} {\bibinfo {author} {\bibnamefont {{R Core Team}}},\
  }\href {https://www.R-project.org/} {\emph {\bibinfo {title} {R: A language
  and evironment for statistical computing}}},\ \bibinfo {organization} {R
  Foundation for Statistical Computing},\ \bibinfo {address} {Vienna, Austria}
  (\bibinfo {year} {2015})\BibitemShut {NoStop}%
\bibitem [{\citenamefont {Ihaka}\ and\ \citenamefont
  {Gentleman}(1996)}]{ihak:gent:1996}%
  \BibitemOpen
  \bibfield  {author} {\bibinfo {author} {\bibfnamefont {R.}~\bibnamefont
  {Ihaka}}\ and\ \bibinfo {author} {\bibfnamefont {R.}~\bibnamefont
  {Gentleman}},\ }\href@noop {} {\bibfield  {journal} {\bibinfo  {journal} {J.
  Comput. Graph. Stat.}\ }\textbf {\bibinfo {volume} {5}},\ \bibinfo {pages}
  {299} (\bibinfo {year} {1996})}\BibitemShut {NoStop}%
\bibitem [{\citenamefont {Adler}, \citenamefont {Murdoch},\ and\ \citenamefont
  {{others}}(2016)}]{rgl}%
  \BibitemOpen
  \bibfield  {author} {\bibinfo {author} {\bibfnamefont {D.}~\bibnamefont
  {Adler}}, \bibinfo {author} {\bibfnamefont {D.}~\bibnamefont {Murdoch}}, \
  and\ \bibinfo {author} {\bibnamefont {{others}}},\ }\href
  {http://CRAN.R-project.org/package=rgl} {\emph {\bibinfo {title} {rgl: 3D
  visualization using openGL}}} (\bibinfo {year} {2016}),\ \bibinfo {note} {r
  package version 0.95.1441}\BibitemShut {NoStop}%
\bibitem [{\citenamefont {Nave}, \citenamefont {Eastoe},\ and\ \citenamefont
  {Penfold}(2000)}]{nave2000so}%
  \BibitemOpen
  \bibfield  {author} {\bibinfo {author} {\bibfnamefont {S.}~\bibnamefont
  {Nave}}, \bibinfo {author} {\bibfnamefont {J.}~\bibnamefont {Eastoe}}, \ and\
  \bibinfo {author} {\bibfnamefont {J.}~\bibnamefont {Penfold}},\ }\href@noop
  {} {\bibfield  {journal} {\bibinfo  {journal} {Langmuir}\ }\textbf {\bibinfo
  {volume} {16}},\ \bibinfo {pages} {8733} (\bibinfo {year}
  {2000})}\BibitemShut {NoStop}%
\bibitem [{\citenamefont {Martin}\ and\ \citenamefont
  {Magid}(1981)}]{martin1981carbon}%
  \BibitemOpen
  \bibfield  {author} {\bibinfo {author} {\bibfnamefont {C.~A.}\ \bibnamefont
  {Martin}}\ and\ \bibinfo {author} {\bibfnamefont {L.~J.}\ \bibnamefont
  {Magid}},\ }\href@noop {} {\bibfield  {journal} {\bibinfo  {journal} {J.
  Phys. Chem.}\ }\textbf {\bibinfo {volume} {85}},\ \bibinfo {pages} {3938}
  (\bibinfo {year} {1981})}\BibitemShut {NoStop}%
\bibitem [{\citenamefont {Hsu}, \citenamefont {Dufresne},\ and\ \citenamefont
  {Weitz}(2005)}]{hsu2005charge}%
  \BibitemOpen
  \bibfield  {author} {\bibinfo {author} {\bibfnamefont {M.~F.}\ \bibnamefont
  {Hsu}}, \bibinfo {author} {\bibfnamefont {E.~R.}\ \bibnamefont {Dufresne}}, \
  and\ \bibinfo {author} {\bibfnamefont {D.~A.}\ \bibnamefont {Weitz}},\
  }\href@noop {} {\bibfield  {journal} {\bibinfo  {journal} {Langmuir}\
  }\textbf {\bibinfo {volume} {21}},\ \bibinfo {pages} {4881} (\bibinfo {year}
  {2005})}\BibitemShut {NoStop}%
\bibitem [{\citenamefont {Briscoe}\ and\ \citenamefont
  {Horn}(2002)}]{briscoe2002direct}%
  \BibitemOpen
  \bibfield  {author} {\bibinfo {author} {\bibfnamefont {W.~H.}\ \bibnamefont
  {Briscoe}}\ and\ \bibinfo {author} {\bibfnamefont {R.~G.}\ \bibnamefont
  {Horn}},\ }\href@noop {} {\bibfield  {journal} {\bibinfo  {journal}
  {Langmuir}\ }\textbf {\bibinfo {volume} {18}},\ \bibinfo {pages} {3945}
  (\bibinfo {year} {2002})}\BibitemShut {NoStop}%
\bibitem [{\citenamefont {Dufresne}, \citenamefont {Hsu},\ and\ \citenamefont
  {Weitz}(2005)}]{dufresne2005reverse}%
  \BibitemOpen
  \bibfield  {author} {\bibinfo {author} {\bibfnamefont {E.~R.}\ \bibnamefont
  {Dufresne}}, \bibinfo {author} {\bibfnamefont {M.~F.}\ \bibnamefont {Hsu}}, \
  and\ \bibinfo {author} {\bibfnamefont {D.~A.}\ \bibnamefont {Weitz}},\
  }\href@noop {} {\bibfield  {journal} {\bibinfo  {journal} {Bull. Am. Phys.
  Soc.}\ } (\bibinfo {year} {2005})}\BibitemShut {NoStop}%
\bibitem [{\citenamefont {Pinero}, \citenamefont {Bhuiyan},\ and\ \citenamefont
  {Bratko}(2004)}]{pinero2004electrostatic}%
  \BibitemOpen
  \bibfield  {author} {\bibinfo {author} {\bibfnamefont {J.}~\bibnamefont
  {Pinero}}, \bibinfo {author} {\bibfnamefont {L.}~\bibnamefont {Bhuiyan}}, \
  and\ \bibinfo {author} {\bibfnamefont {D.}~\bibnamefont {Bratko}},\
  }\href@noop {} {\bibfield  {journal} {\bibinfo  {journal} {J. Chem. Phys.}\
  }\textbf {\bibinfo {volume} {120}},\ \bibinfo {pages} {11941} (\bibinfo
  {year} {2004})}\BibitemShut {NoStop}%
\bibitem [{\citenamefont {Roberts}\ \emph {et~al.}(2008)\citenamefont
  {Roberts}, \citenamefont {Sanchez}, \citenamefont {Kemp}, \citenamefont
  {Wood},\ and\ \citenamefont {Bartlett}}]{roberts2008electrostatic}%
  \BibitemOpen
  \bibfield  {author} {\bibinfo {author} {\bibfnamefont {G.~S.}\ \bibnamefont
  {Roberts}}, \bibinfo {author} {\bibfnamefont {R.}~\bibnamefont {Sanchez}},
  \bibinfo {author} {\bibfnamefont {R.}~\bibnamefont {Kemp}}, \bibinfo {author}
  {\bibfnamefont {T.}~\bibnamefont {Wood}}, \ and\ \bibinfo {author}
  {\bibfnamefont {P.}~\bibnamefont {Bartlett}},\ }\href@noop {} {\bibfield
  {journal} {\bibinfo  {journal} {Langmuir}\ }\textbf {\bibinfo {volume}
  {24}},\ \bibinfo {pages} {6530} (\bibinfo {year} {2008})}\BibitemShut
  {NoStop}%
\bibitem [{\citenamefont {Sainis}\ \emph {et~al.}(2008)\citenamefont {Sainis},
  \citenamefont {Germain}, \citenamefont {Mejean},\ and\ \citenamefont
  {Dufresne}}]{sainis2008electrostatic}%
  \BibitemOpen
  \bibfield  {author} {\bibinfo {author} {\bibfnamefont {S.~K.}\ \bibnamefont
  {Sainis}}, \bibinfo {author} {\bibfnamefont {V.}~\bibnamefont {Germain}},
  \bibinfo {author} {\bibfnamefont {C.~O.}\ \bibnamefont {Mejean}}, \ and\
  \bibinfo {author} {\bibfnamefont {E.~R.}\ \bibnamefont {Dufresne}},\
  }\href@noop {} {\bibfield  {journal} {\bibinfo  {journal} {Langmuir}\
  }\textbf {\bibinfo {volume} {24}},\ \bibinfo {pages} {1160} (\bibinfo {year}
  {2008})}\BibitemShut {NoStop}%
\bibitem [{\citenamefont {Eicke}, \citenamefont {Borkovec},\ and\ \citenamefont
  {Das-Gupta}(1989)}]{eicke1989conductivity}%
  \BibitemOpen
  \bibfield  {author} {\bibinfo {author} {\bibfnamefont {H.~F.}\ \bibnamefont
  {Eicke}}, \bibinfo {author} {\bibfnamefont {M.}~\bibnamefont {Borkovec}}, \
  and\ \bibinfo {author} {\bibfnamefont {B.}~\bibnamefont {Das-Gupta}},\
  }\href@noop {} {\bibfield  {journal} {\bibinfo  {journal} {J. Phys. Chem.}\
  }\textbf {\bibinfo {volume} {93}},\ \bibinfo {pages} {314} (\bibinfo {year}
  {1989})}\BibitemShut {NoStop}%
\bibitem [{\citenamefont {De}\ and\ \citenamefont
  {Maitra}(1995)}]{de_solution_1995}%
  \BibitemOpen
  \bibfield  {author} {\bibinfo {author} {\bibfnamefont {T.~K.}\ \bibnamefont
  {De}}\ and\ \bibinfo {author} {\bibfnamefont {A.}~\bibnamefont {Maitra}},\
  }\href {\doibase 10.1016/0001-8686(95)80005-N} {\bibfield  {journal}
  {\bibinfo  {journal} {Advances in Colloid and Interface Science}\ }\textbf
  {\bibinfo {volume} {59}},\ \bibinfo {pages} {95} (\bibinfo {year}
  {1995})}\BibitemShut {NoStop}%
\bibitem [{\citenamefont {Mathew}\ \emph {et~al.}(1988)\citenamefont {Mathew},
  \citenamefont {Patanjali}, \citenamefont {Nabi},\ and\ \citenamefont
  {Maitra}}]{mathew_concept_1988}%
  \BibitemOpen
  \bibfield  {author} {\bibinfo {author} {\bibfnamefont {C.}~\bibnamefont
  {Mathew}}, \bibinfo {author} {\bibfnamefont {P.~K.}\ \bibnamefont
  {Patanjali}}, \bibinfo {author} {\bibfnamefont {A.}~\bibnamefont {Nabi}}, \
  and\ \bibinfo {author} {\bibfnamefont {A.}~\bibnamefont {Maitra}},\ }\href
  {\doibase 10.1016/0166-6622(88)80128-8} {\bibfield  {journal} {\bibinfo
  {journal} {Colloids and Surfaces}\ }\textbf {\bibinfo {volume} {30}},\
  \bibinfo {pages} {253} (\bibinfo {year} {1988})}\BibitemShut {NoStop}%
\bibitem [{\citenamefont {Grest}\ \emph {et~al.}(1986)\citenamefont {Grest},
  \citenamefont {Webman}, \citenamefont {Safran},\ and\ \citenamefont
  {Bug}}]{grest_dynamic_1986}%
  \BibitemOpen
  \bibfield  {author} {\bibinfo {author} {\bibfnamefont {G.~S.}\ \bibnamefont
  {Grest}}, \bibinfo {author} {\bibfnamefont {I.}~\bibnamefont {Webman}},
  \bibinfo {author} {\bibfnamefont {S.~A.}\ \bibnamefont {Safran}}, \ and\
  \bibinfo {author} {\bibfnamefont {A.~L.~R.}\ \bibnamefont {Bug}},\ }\href
  {\doibase 10.1103/PhysRevA.33.2842} {\bibfield  {journal} {\bibinfo
  {journal} {Phys. Rev. A}\ }\textbf {\bibinfo {volume} {33}},\ \bibinfo
  {pages} {2842} (\bibinfo {year} {1986})}\BibitemShut {NoStop}%
\bibitem [{\citenamefont {Huang}\ \emph {et~al.}(1984)\citenamefont {Huang},
  \citenamefont {Safran}, \citenamefont {Kim}, \citenamefont {Grest},
  \citenamefont {Kotlarchyk},\ and\ \citenamefont
  {Quirke}}]{huang_attractive_1984}%
  \BibitemOpen
  \bibfield  {author} {\bibinfo {author} {\bibfnamefont {J.~S.}\ \bibnamefont
  {Huang}}, \bibinfo {author} {\bibfnamefont {S.~A.}\ \bibnamefont {Safran}},
  \bibinfo {author} {\bibfnamefont {M.~W.}\ \bibnamefont {Kim}}, \bibinfo
  {author} {\bibfnamefont {G.~S.}\ \bibnamefont {Grest}}, \bibinfo {author}
  {\bibfnamefont {M.}~\bibnamefont {Kotlarchyk}}, \ and\ \bibinfo {author}
  {\bibfnamefont {N.}~\bibnamefont {Quirke}},\ }\href {\doibase
  10.1103/PhysRevLett.53.592} {\bibfield  {journal} {\bibinfo  {journal} {Phys.
  Rev. Lett.}\ }\textbf {\bibinfo {volume} {53}},\ \bibinfo {pages} {592}
  (\bibinfo {year} {1984})}\BibitemShut {NoStop}%
\bibitem [{\citenamefont {Diamant}\ and\ \citenamefont
  {Andelman}(2016)}]{diamant2016free}%
  \BibitemOpen
  \bibfield  {author} {\bibinfo {author} {\bibfnamefont {H.}~\bibnamefont
  {Diamant}}\ and\ \bibinfo {author} {\bibfnamefont {D.}~\bibnamefont
  {Andelman}},\ }\href@noop {} {\bibfield  {journal} {\bibinfo  {journal}
  {Curr. Opin. Colloid Interface Sci.}\ }\textbf {\bibinfo {volume} {22}},\
  \bibinfo {pages} {94} (\bibinfo {year} {2016})}\BibitemShut {NoStop}%
\bibitem [{\citenamefont {Eicke}(1980)}]{eicke1980surfactants}%
  \BibitemOpen
  \bibfield  {author} {\bibinfo {author} {\bibfnamefont {H.-F.}\ \bibnamefont
  {Eicke}},\ }in\ \href@noop {} {\emph {\bibinfo {booktitle} {Micelles}}}\
  (\bibinfo  {publisher} {Springer},\ \bibinfo {year} {1980})\ pp.\ \bibinfo
  {pages} {85--145}\BibitemShut {NoStop}%
\bibitem [{\citenamefont {Tamamushi}\ and\ \citenamefont
  {Watanabe}(1980)}]{tamamushi1980formation}%
  \BibitemOpen
  \bibfield  {author} {\bibinfo {author} {\bibfnamefont {B.}~\bibnamefont
  {Tamamushi}}\ and\ \bibinfo {author} {\bibfnamefont {N.}~\bibnamefont
  {Watanabe}},\ }\href@noop {} {\bibfield  {journal} {\bibinfo  {journal}
  {Colloid \& Polym. Sci.}\ }\textbf {\bibinfo {volume} {258}},\ \bibinfo
  {pages} {174} (\bibinfo {year} {1980})}\BibitemShut {NoStop}%
\bibitem [{\citenamefont {Kislenko}\ and\ \citenamefont
  {Razumov}(2017)}]{kislenko2017molecular}%
  \BibitemOpen
  \bibfield  {author} {\bibinfo {author} {\bibfnamefont {S.}~\bibnamefont
  {Kislenko}}\ and\ \bibinfo {author} {\bibfnamefont {V.}~\bibnamefont
  {Razumov}},\ }\href@noop {} {\bibfield  {journal} {\bibinfo  {journal}
  {Colloid J.}\ }\textbf {\bibinfo {volume} {79}},\ \bibinfo {pages} {76}
  (\bibinfo {year} {2017})}\BibitemShut {NoStop}%
\bibitem [{\citenamefont {Fletcher}, \citenamefont {Howe},\ and\ \citenamefont
  {Robinson}(1987)}]{fletcher1987kinetics}%
  \BibitemOpen
  \bibfield  {author} {\bibinfo {author} {\bibfnamefont {P.~D.}\ \bibnamefont
  {Fletcher}}, \bibinfo {author} {\bibfnamefont {A.~M.}\ \bibnamefont {Howe}},
  \ and\ \bibinfo {author} {\bibfnamefont {B.~H.}\ \bibnamefont {Robinson}},\
  }\href@noop {} {\bibfield  {journal} {\bibinfo  {journal} {J. Chem. Soc.,
  Fara. Trans. 1: Phys. Chem. Cond. Phases}\ }\textbf {\bibinfo {volume}
  {83}},\ \bibinfo {pages} {985} (\bibinfo {year} {1987})}\BibitemShut
  {NoStop}%
\bibitem [{\citenamefont {Mejuto}\ \emph {et~al.}(2014)\citenamefont {Mejuto},
  \citenamefont {Morales}, \citenamefont {Moldes},\ and\ \citenamefont
  {Cid}}]{mejuto2014effects}%
  \BibitemOpen
  \bibfield  {author} {\bibinfo {author} {\bibfnamefont {J.}~\bibnamefont
  {Mejuto}}, \bibinfo {author} {\bibfnamefont {J.}~\bibnamefont {Morales}},
  \bibinfo {author} {\bibfnamefont {O.}~\bibnamefont {Moldes}}, \ and\ \bibinfo
  {author} {\bibfnamefont {A.}~\bibnamefont {Cid}},\ }\href@noop {} {\bibfield
  {journal} {\bibinfo  {journal} {J. Appl. Solution Chem. Model.}\ }\textbf
  {\bibinfo {volume} {3}},\ \bibinfo {pages} {106} (\bibinfo {year}
  {2014})}\BibitemShut {NoStop}%
\end{thebibliography}%

\newpage

\begin{table*}[htb]
\caption{\label{compositions}
Solution composition and average radius of 
gyration ($R_g$), asphericity ($A_s$), and chemical potential for growth to an ($n$+1)-mer}
\centering
\begin{tabular}{ c c | c c c }
N$_{\text{AOT}}$ & N$_{\text{ISO}}$ & $R_g [nm]$ & $A_s$ & $\Delta \mu_{n+1}^{0}$ [kcal/mol] \\
\hline
1   & 3500 & N/A & N/A & -13.44 $\pm$ 0.47 \\
10  & 3500 & 1.04 $\pm$ 0.01 & 0.48 $\pm$ 0.002 & -55.9 $\pm$ 2.4 \\
20  & 3500 & 1.17 $\pm$ 0.01 & 0.18 $\pm$ 0.001 & -89.0 $\pm$ 4.4 \\
30  & 3500 & 1.31 $\pm$ 0.28 & 0.21 $\pm$ 0.001 & -78.2 $\pm$ 3.7 \\
40  & 3500 & 1.47 $\pm$ 0.01 & 0.16 $\pm$ 0.008 & -85.0 $\pm$ 5.3 \\
50  & 3500 & 1.58 $\pm$ 0.01 & 0.18 $\pm$ 0.001 & -84.1 $\pm$ 5.1 \\
60  & 3500 & 1.67 $\pm$ 0.01 & 0.11 $\pm$ 0.001 & -95.9 $\pm$ 5.9 \\
70  & 3500 & 1.79 $\pm$ 0.01 & 0.20 $\pm$ 0.001 & -93.0 $\pm$ 4.5 \\
80  & 7000 & 1.78 $\pm$ 0.01 & 0.22 $\pm$ 0.001 & -90.5 $\pm$ 10.3 \\
90  & 7000 & 1.97 $\pm$ 0.02 & 0.18 $\pm$ 0.001 & -91.3 $\pm$ 12.6 \\
100 & 7000 & 2.07 $\pm$ 0.01 & 0.10 $\pm$ 0.001 & -85.2 $\pm$ 9.3
\end{tabular}
\end{table*}

\clearpage
\newpage

\listoffigures


\end{document}